\documentclass[%
 reprint,
 amsmath,amssymb,
 aps,
]{revtex4-2}
\setcitestyle{super}
\usepackage{hyperref}
\usepackage{graphicx}
\usepackage{dcolumn}
\usepackage{bm}
\usepackage{xcolor}
\usepackage{color}

\begin{document}

\title{Tuning the Topological Properties of the Antiferromagnetic \\ V(Bi$_{1-x}$Sb$_{x}$)$_{2}$Te$_{4}$ via Sb concentration
}

\author{D. A-León}
\email{daleon@unal.edu.co}
\author{D.A. Landínez Téllez}%
\author{J. Roa-Rojas}%
\affiliation{%
Grupo de Física de Nuevos Materiales, Departamento de Física, Universidad Nacional de Colombia, Bogotá, Colombia
}

\author{Rafael González-Hernández}%
\email{rhernandezj@uninorte.edu.co}
\affiliation{Departamento de F\'{i}sica, Universidad del Valle, Cali, Colombia}
\affiliation{Departamento de F\'{i}sica y Geociencias, Universidad del Norte, Barranquilla, Colombia}

\date{\today}

\begin{abstract}  
	The investigation of topological materials has uncovered groundbreaking phases of matter with significant implications for quantum technologies. Here, we explore the antiferromagnetic topological insulator family V(Bi$_{1-x}$Sb$_{x}$)$_{2}$Te$_{4}$ ($x$=$0$, $0.5$, $1$), formed by introducing vanadium telluride (VTe) layers into the layered topological insulator (Bi$_{1-x}$Sb$_{x}$)$_{2}$Te$_{3}$. Our results reveal the tunability of the spin Hall conductivity (SHC) and its topological contribution, quantified by the recently introduced average Spin Chern Number (ASCN), via Sb concentration. The materials' strong topological insulating behavior is established through spin-orbit coupling-induced band inversions, nontrivial $\mathbb{Z}_2$ invariants, and the presence of topological surface states. These findings position V(Bi$_{1-x}$Sb$_{x}$)$_{2}$Te$_{4}$ as promising candidates for next-generation spintronic devices and advanced quantum applications.  
\end{abstract} 

\maketitle

\section{\label{sec:level1}Introduction}

Insulators are a class of materials defined by the absence of conduction states, resulting from a band gap separating the valence and conduction bands. However, topological insulators represent an intriguing exception to this rule. While their bulk remains insulating, their surfaces host conductive channels that are protected by crystal symmetries, rendering them robust against external perturbations~\cite{qi2011topological,kumar2020topological,xiao2021first,cava2013crystal}. 
This exceptional property has sparked considerable interest in the scientific community, as it paves the way for investigating exotic phases of ma\-tter, including the topological insulators, quantum spin Hall insulators, Dirac and Weyl semimetals, and other novel phenomena~\cite{yu2010quantized,zou2017observation,konig2007quantum,yan2017topological}.

The bulk boundary correspondence is a fundamental principle used to identify non-trivial topological materials by bulk invariants. This principle confirms the existence of conducting states at the boundaries of such materials and has been instrumental in the discovery of topological insulators such as Bi$_{2}$Te$_{3}$, Bi$_{2}$Sb$_{3}$, and (Bi$_{1-}$Sb$_{x}$)$_{2}$Te$_{3}$~\cite{zhang2009topological,zhang2010first,mazumder2021brief,he2013review}. These materials exhibit exotic conduction properties that can only be explained by their topological nature. 
Later, researchers have extended this field by exploring materials that combine these exotic properties with intrinsic magnetic states, driven by their potential applications in advanced technological devices. This new line of research has led to the development of magnetic topological insulators (MTIs). The first step in this direction was the doping of thin films with magnetic ions such as vanadium (V). Notable in this field is the work of Chang et al~\cite{chang2015high}, who report the creation of the quantum anomalous state in ultra thin films of (Bi,Sb)Te$_{3}$ doped with V and deposited on a SrTiO$_{3}$ substrate. Over time, a breakthrough occurred in 2019 with the theoretical prediction and experimental realization of the first MTI, MnBi$_{2}$Te$_{4}$~\cite{otrokov2019prediction,wu2019natural}. This material, formed by incorporating 3$d$ transition metals into the well-known topological insulator Bi$_{2}$Te$_{3}$, has attracted significant attention due to its interplay between magnetism and topology. Subsequent studies explored Mn-based variants, including MnSb$_{2}$Te$_{4}$~\cite{eremeev2017competing,chen2020electronic,zhou2020topological} and Mn(Bi$_{1-x}$Sb$_{x}$)$_2$Te$_4$~\cite{lei2020magnetized,lee2021evidence,yan2019evolution}, demonstrating that antiferromagnetic topological properties persist across these systems. More recently, V-based analogs such as VBi$_{2}$Te$_{4}$~\cite{li2019intrinsic,petrov2021domain,zhang2023strain,altena2023phase} have been investigated from first-principles calculations, with experimental validation yet to be realized.

In this study, we examine the structural and magnetic stability of VTe layers integrated into the well-defined atomic architecture of the van der Waals topological insulator (Bi$_{1-x}$Sb$_{x}$)$_{2}$Te$_{3}$.
This modification facilitates the exploration of stoichiometrically consistent bismuth and antimony layer arrangements, providing a framework to examine the physical properties of V(Bi$_{1-x}$Sb$_{x}$)$_{2}$Te$_{4}$. 
We focus on the structural characteristics of the double-unit-cell configurations, highlighting the impact of magnetic configurations on the material's topological properties. 
Our results provide clear evidence of band inversion, co\-rroborated by surface-state calculations and the eva\-luation of the topological invariant $\mathbb{Z}_{2}$. Importantly, we demonstrate that V(Bi$_{1-x}$Sb$_{x}$)$_{2}$Te$_{4}$ crystals with $x = 0$, $x = 0.5$, and $x = 1$ exhibit magnetic topological insulating (MTI) behavior. 
Furthermore, we employ the recently developed definition of the average spin Chern number (ASCN) \cite{PhysRevB.110.125129} to quantify the topological response to the spin Hall conductivity (SHC) and to tune the topological properties of V(Bi$_{1-x}$Sb$_{x}$)$_{2}$Te$_{4}$ by varying the Sb concentration. 
This approach represents a significant advance in the discovery of new MTIs and the controlled tuning of their topological properties, contributing to a broader understanding of topological materials with antiferromagnetic phases.

\section{\label{sec:level22}Methods}

The $ab$-$initio$ calculations were conducted within the framework of Density Functional Theory (DFT) using the Vienna Ab-Initio Simulation Package (VASP)~\cite{vasp}. The projector augmented wave (PAW) method was applied, alongside the generalized gradient approximation (GGA) for the exchange-correlation functional. To ensure accu\-rate results, convergence criteria were set to an energy precision of $1.0$$\times$$10^{-6}$ eV and a force tolerance of 0.01 eV/$\textup{\AA}$.
For structural characterization, atomic positions were initialized based on the double rhombohedral or double hexagonal cell configurations reported by \mbox{Eremeev} et al.~\cite{eremeev2017competing}. During structural optimization, both atomic positions and unit cell geometries were allowed to relax. 
To determine the appropriate energy cutoff for the plane-wave expansion, test optimization calculations were performed. These tests confirmed that an energy cutoff of 450 eV, as suggested by Li et al.~\cite{li2022}, was sufficient for V(Bi$_{1-x}$Sb$_{x}$)$_{2}$Te$_{4}$ crystals with $x = 0$, $x = 0.5$, and $x = 1$.  
The Hubbard-Dudarev correction~\cite{dudarev1998} was \mbox{applied} to the $d$-V states, with $U = 3.25$ eV and $J = 0$ eV, in agreement with the values reported in \cite{cococcioni2005linear,zhou2004first,wang2004enthalpy}.
Due to the layered nature of the material, van der Waals in\-teractions were incorporated using Grimme’s method with a zero-damping function~\cite{grimme2010consistent}. The first Brillouin zone was sampled using a $9$$\times$$9$$\times$3 $k$-point mesh to ensure accurate integration in reciprocal space.

For the calculation of electronic and magnetic properties, we use the atomic positions and lattice parameters of the most energetically stable configurations. Specifically, rhombohedral cells are employed for $x$=0 and $x$=1, while hexagonal cells are used for $x$=0.5.
For the determination of topological properties,  IrRep code was used to compute the topological invariant $Z_2$ by counting the number of Kramers pairs with odd-parity eigenvalues at the time-reversal invariant momenta (TRIM) points, as described in Refs.~\cite{khalaf2018symmetry,song2018quantitative,tanaka2020theory,wang2019higher}.
Finally, \mbox{using} maximally localized Wannier functions (MLWFs) \cite{mostofi2008wannier90}, tight-binding models were used to accurately reproduce the electronic band structure, \mbox{including} SOC, within an energy range of $E_F$ $\mp$ 2 eV, and con\-sidering the $d$-orbitals of V and the $p$-orbitals of Bi, Te, and Sb. 
The surface spectral function and spin Hall conductivity (SHC) were calculated using the surface Green’s function method and the Kubo formula, respectively, as implemented in WannierTools code\cite{WU2017}. 
For the surface states, a 21-layer slab model was used, while a dense $160^3$ $k$-point mesh was employed for the SHC calculations to ensure high accuracy in the transport properties.
The average spin Chern number (ASCN) was computed according to the procedure described in Ref \cite{PhysRevB.110.125129}.


\section{\label{sec:level33}Results and discussion}

\begin{figure}[h]
	\includegraphics[width=0.42\textwidth]{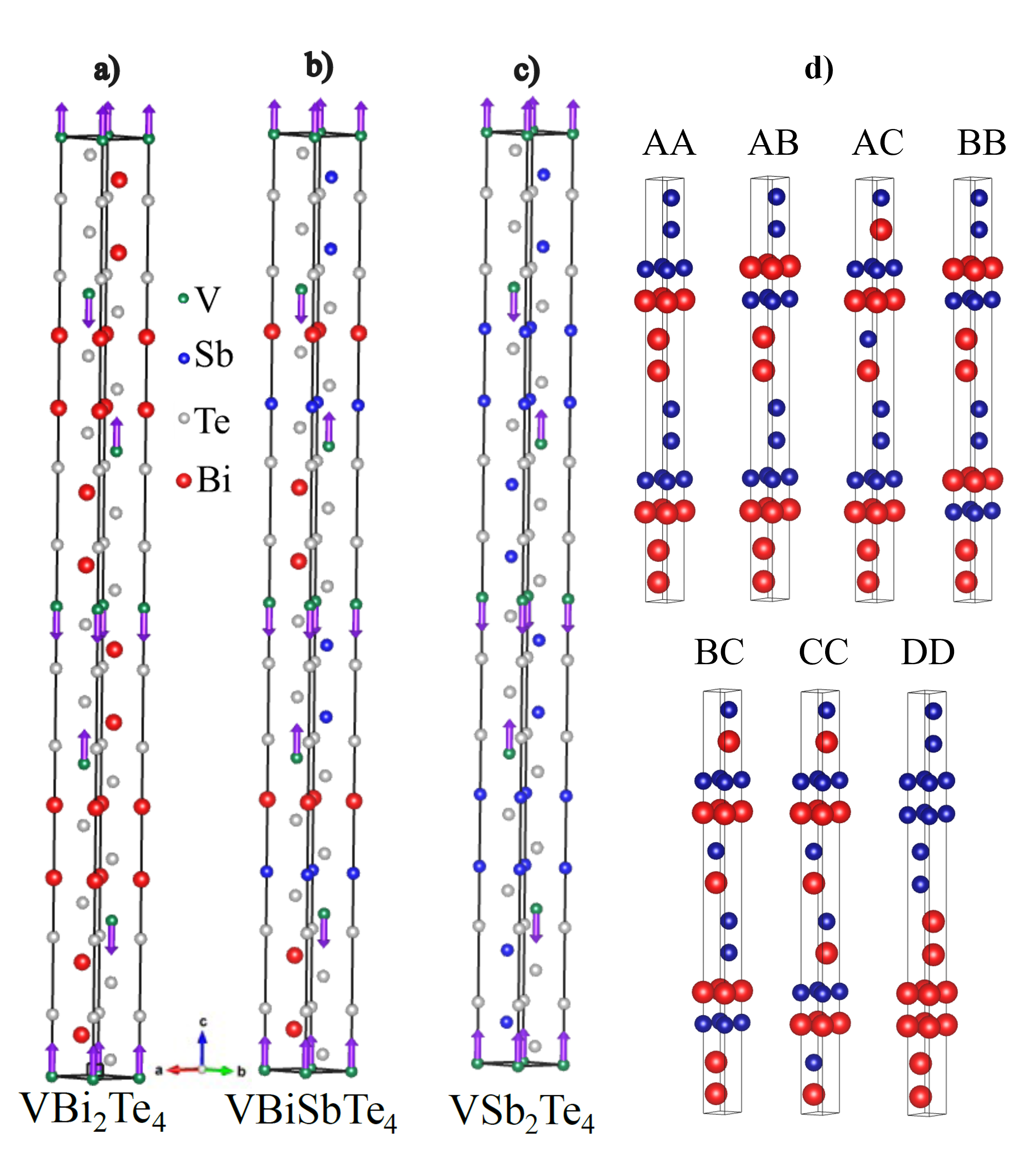}
	\caption{\label{figcristales} Ball-and-stick representation of the crystal structure of V(Bi$_{1-x}$Sb$_{x}$)$_{2}$Te$_{4}$ for $x$=$0$, $x$=$0.5$, and $x$=$1$, obtained from structural relaxation without SOC. Vanadium atoms are depicted as green spheres, bismuth atoms as red spheres, antimony atoms as blue spheres, and tellurium atoms as gray spheres. The purple arrows represent the magnetic moment of the V atoms within each layer. a) shows the hexagonal structure of VBi$_{2}$Te$_{4}$ with the AFM2 configuration, b) illustrates the most stable AFM BB configuration for the VBiSbTe$_{4}$ crystal, c) provides a view of AFM2 hexagonal structure of VSb$_{2}$Te$_{4}$, and d) details the di\-fferent layer configurations of bismuth and antimony in the VBiSbTe$_{4}$ structure.}
\end{figure}

For V(Bi$_{1-x}$Sb$_{x}$)$_{2}$Te$_{4}$ with $x$=0 and $x$=1, the materials crystallize in space group \#166, with septuple layers arranged as Te1-Bi(Sb)-Te2-V-Te2-Bi(Sb)-Te. 
Figure \ref{figcristales} illustrates the layered structure of these crystals within their hexagonal unit cell. This structure can be charac\-terized by two possible unit cell configurations: one rhombohedral and one hexagonal. In both cases, the number of V atoms per unit cell is odd. To accommodate antiferromagnetic (AFM) characteristics, the hexagonal (rhombohedral) unit cell must be doubled along the $c$-axis, as described in Ref \cite{otrokov2019prediction}. 

We present the crystal structures of VBi$_{2}$Te$_{4}$ in Fig. \ref{figcristales}a and VSb$_{2}$Te$_{4}$ in Fig. \ref{figcristales}c. For this study, three distinct magnetic configurations were explored for each crystal. The first is a ferromagnetic (FM) configuration, where the magnetic moments of all V atoms are aligned along the $c$-axis. The second configuration denoted AFM1, represents a triple-layer antiferromagnetic arrangement, where the magnetic moments of the first three V layers point along the $c$-axis, while the magnetic moments of the subsequent three layers point in the opposite direction, along -$c$. The third configuration (AFM2), shown in Fig. \ref{figcristales}, is a layer-coupled AFM structure (AFM type A), where the magnetic moments in each plane are aligned in the same direction, alternating between $c$ and -$c$ across layers.

The energy minimization calculations reveal that the AFM2 structure is the most stable for both VBi$_{2}$Te$_{4}$ and VSb$_{2}$Te$_{4}$ crystals, as shown in Figures \ref{figcristales}a and \ref{figcristales}c. 
This result aligns with previous studies on the AFM2 phase in MnBi${2}$Te${4}$~\cite{otrokov2019prediction,yan2019crystal,zhou2020topological}, which report similar stability trends.
These materials crystallize in space group  \#166 (R-3m), corresponding to a rhombohedral phase, where the three lattice vectors have equal length and form non-orthogonal angles.

In the hexagonal form, VBi$_{2}$Te$_{4}$ exhibits lattice para\-meters $a$=$b$=4.3539$\,\textup{\AA}$ and $c$=80.9494$\,\textup{\AA}$, with a unit cell volume of 1328.95 $\textup{\AA}^3$. 
This structure includes inversion symmetry ($\mathbb{P}$), a key feature in the calculation of the $\mathbb{Z}_2$ index. Similarly, the hexagonal unit cell for VSb$_{2}$Te$_{4}$ has lattice parameters $a$=$b$=4.2793$\,\textup{\AA}$ and $c$=80.2408$\,\textup{\AA}$, with a unit cell volume of 1272.57 $\textup{\AA}^3$, in agreement with previously reported values~\cite{li2019intrinsic,zhang2023strain}. 
The structural charac\-teristics, magnetic configu\-rations, lattice constants ($a$ and $c$), and energy comparisons with respect to the most stable configu\-ration are summarized in Table \ref{tab:table101}. This table highlights that certain AFM orderings impose unique symmetry constraints, with electronic states protected by the combined symmetry of partial translation and time-reversal ($t_{1/2}\mathbb{T}$).

\begin{table}[htb]
\caption{\label{tab:table101}%
Structural properties of V(Bi$_{1-x}$Sb$_{x}$)$_{2}$Te$_{4}$ crystals with $x$=0, $x$=0.5, and $x$=1, for different magnetic configurations, obtained from structural relaxation without SOC. The notation for the magnetic configuration of VBiSbTe$_{4}$ corresponds to the layered distributions of bismuth and antimony as shown in Figure \ref{figcristales}. Relevant magnetic symme\-tries present are also specified, with $\mathbb{P}$ indicating inversion symmetry and $t_{1/2}\mathbb{T}$ representing the combined partial-translation and time-reversal symmetry.}
\begin{ruledtabular}
\begin{tabular}{ l c c c c c c}
\vspace{1mm}
Material  &Magnetic  & $a(\textup{\r{A}})$ & $c(\textup{\r{A}})$ &   $\Delta E_{T}$& Sym.   \\ 
  & config. &  &    &  (meV)  \\ 
\hline
VBi$_{2}$Te$_{4}$ & AFM2 & 4.353  & 80.949   &0.00 & $\mathbb{P}$,$t_{1/2}\mathbb{T}$\\
VBi$_{2}$Te$_{4}$ & AFM1 &4.351 & 81.235  &0.46  & $\mathbb{P}$,$t_{1/2}\mathbb{T}$  \\
VBi$_{2}$Te$_{4}$ & FM & 4.353 & 81.095  &1.62 & $\mathbb{P}$  \\
VBi$_{2}$Te$_{4}$\footnotemark[1] & - & 4.370 & 81.200  &-\\
VBi$_{2}$Te$_{4}$\footnotemark[2]& - & 4.350  & 82.620  &- \\
VBi$_{2}$Te$_{4}$\footnotemark[3]& - & 4.338  & 80.344  &- \\
\hline
VBiSbTe$_{4}$ & AFM2 BB & 4.318&80.654 &0.0&$t_{1/2}\mathbb{T}$ \\
VBiSbTe$_{4}$ & AFM1 BB & 4.317&80.804 &0.82&$t_{1/2}\mathbb{T}$\\
VBiSbTe$_{4}$ & AFM2 AC &4.317&80.855 &2.74&\\
VBiSbTe$_{4}$ & AFM1 AA &4.317&80.765 &3.13&$t_{1/2}\mathbb{T}$\\
VBiSbTe$_{4}$ & AFM2 CC &4.318&80.624 &3.32&$t_{1/2}\mathbb{T}$\\
VBiSbTe$_{4}$ & AFM2 AA &4.316&80.892 &3.4&$t_{1/2}\mathbb{T}$\\
VBiSbTe$_{4}$ & AFM1 CC &4.317&80.892 &4.93&$t_{1/2}\mathbb{T}$\\
VBiSbTe$_{4}$ & AFM2 BC &4.318&80.730 &4.94\\
VBiSbTe$_{4}$ & AFM1 AC &4.316&80.839 &5.83\\
VBiSbTe$_{4}$ & AFM1 BC &4.317&80.692 &6.83\\
VBiSbTe$_{4}$ & AFM2 AB &4.317&80.768 &7.32\\
VBiSbTe$_{4}$ & AFM1 AB &4.317&80.774 &8.77\\
VBiSbTe$_{4}$ & AFM1 DD &4.315&80.976 &18.91\\
VBiSbTe$_{4}$ & AFM2 DD &4.317&80.883 &27.8&$\mathbb{P}\mathbb{T}$\\\hline
VSb$_{2}$Te$_{4}$ & AFM2 & 4.279 & 80.240 &0.00 & $\mathbb{P}$,$t_{1/2}\mathbb{T}$ \\
VSb$_{2}$Te$_{4}$ & AFM1 &4.281 & 80.216  &2.01 & $\mathbb{P}$,$t_{1/2}\mathbb{T}$\\
VSb$_{2}$Te$_{4}$ & FM & 4.282 & 80.192  & 3.41& $\mathbb{P}$
 \\
VSb$_{2}$Te$_{4}$\footnotemark[3]& - & 4.260 & 77.767  &- \\
\end{tabular}
\end{ruledtabular}
\footnotetext[1]{from Ref.\cite{li2019intrinsic}.}
\footnotetext[2]{from Ref. \cite{zhang2023strain}.}
\footnotetext[2]{from Ref. \cite{petrov2021domain}.}
\end{table}

Total energy calculations of compositions such as V(Bi$_{1-x}$Sb$_{x}$)$_{2}$Te$_{4}$, particularly with $x$=0.5, present a research challenge due to the large number of unit cells that must be considered ~\cite{otrokov2019prediction}. This complexity arises because the layering arrangement must adhere to the stoichiometric formula. This is one reason why there are relatively few $ab$-$initio$ studies on such specific stoichiometric combinations.
Given that the material grows in atomic layers, and considering that the six bismuth and antimony planes within the hexagonal unit cell can be interchanged while maintaining the stoichiometric formula, we identified three base unit cells. The first, which we refer to as Cell A, consists of a triple layer of bismuth followed by a triple layer of antimony. The second, Cell B, is composed of an intercalation of planes in the sequence Bi-Bi-Sb-Bi-Sb-Sb. The third, Cell C, alternates between bismuth and antimony planes along the $c$-axis, as shown in Figure~\ref{figcristales}d.

As previously noted, obtaining AFM configurations requires the use of a double unit cell along the $c$-axis. Given that we have three base unit cells, we proceed by combining them to form the following double unit cells: AA, AB, AC, BB, BC, and CC. Additionally, an alternative cell, designated DD, is identified, in which the first six layers are bismuth and the second six layers are antimony. The organization of these different bismuth and antimony layers along the $c$-axis is depicted in Fi\-gure~\ref{figcristales}d.
Then, we have seven distinct double unit cells for VBiSbTe$_{4}$, each of which was calculated with both AFM1 and AFM2 magnetic configurations. The results of the structural properties for each of these configurations, including magnetic properties, lattice parameters ($a$ and $c$), and the energy differences relative to the most stable configuration, are summarized in Table~\ref{tab:table101}. 

\begin{figure}[t]
\includegraphics[width=0.38\textwidth]{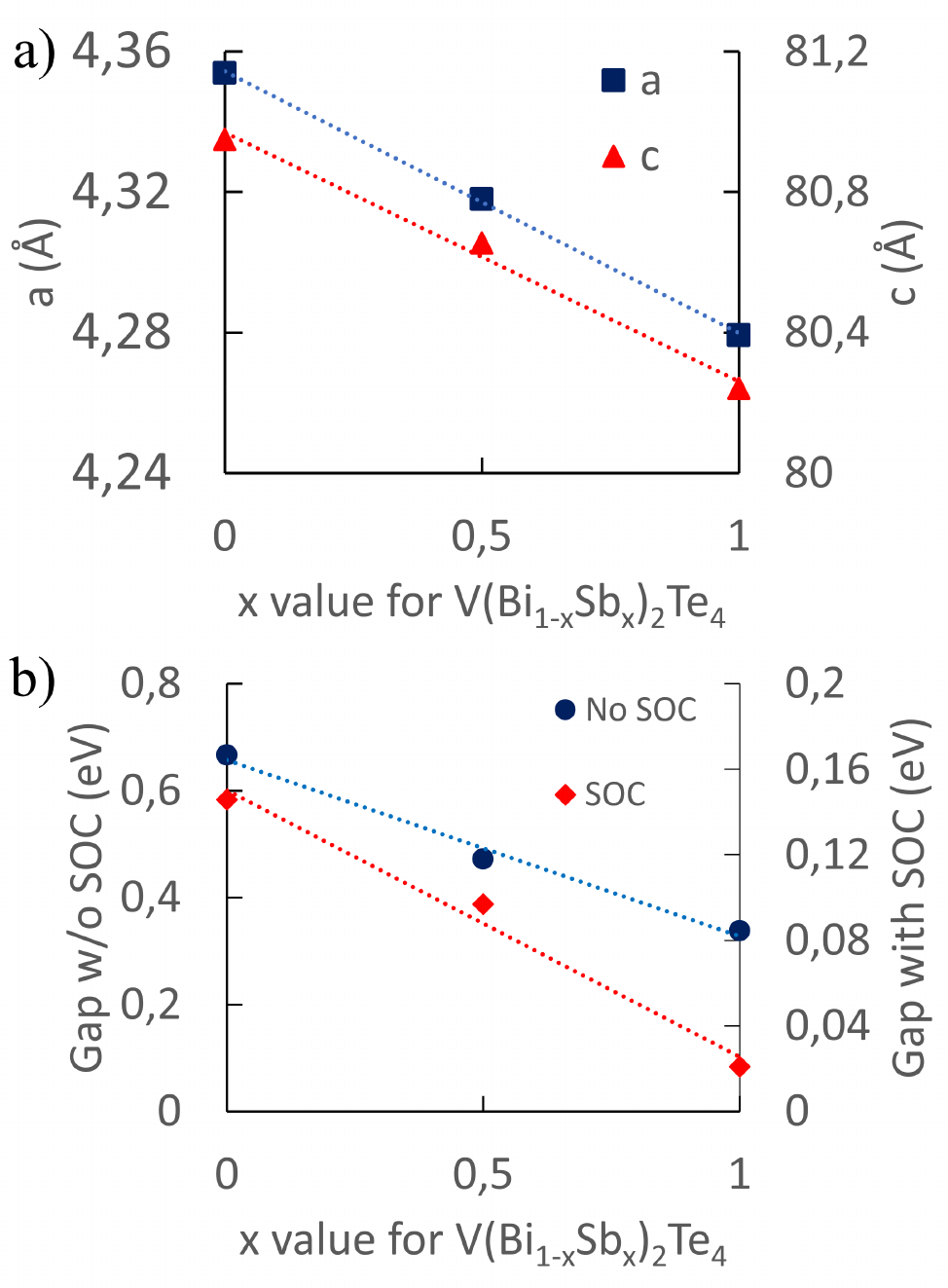}
\caption{\label{lattices}Evolution of lattice parameters $a$ and $c$ in addition to the band gap with the exclusion and inclusion of the spin-orbit coupling for the crystal V(Bi$_{1-x}$Sb$_{x}$)$_{2}$Te$_{4}$ a) presents the relationship between the lattice constants a and c as the concentration of antimony in the crystal increases, a decrease in these parameters is found as the value of $x$ increases. b) presents the direct relationship between the band gap size and the increase of the concentration of Sb atoms in the crystal, it is observed that as the value of $x$ increases, the gaps with both the exclusion and inclusion of the spin-orbit coupling decrease.}
\end{figure}

For the VBiSbTe$_{4}$, we found that the most energetically stable configuration corresponds to the double unit cell with an AFM2 BB-type arrangement, where the V magnetic moments are intercalated along the $c$-axis, as depicted in Figure~\ref{figcristales}b. This double unit cell crystallizes in space group \#156 (P-3m1), a trigonal (or hexagonal) system. This space group exhibits a 120° rotational symmetry operation around the $c$-axis and is hexagonal primi\-tive, with lattice parameters $a$=$b$=4.3180 \AA \  and a perpendicular $c$-axis value of 80.6546 \AA. The volume of this double unit cell is 1302.35 \AA$^3$. Notably, this space group lacks inversion symmetry, as depicted in the atomic configuration shown in Figure~\ref{figcristales}b.

Figure~\ref{lattices}a shows the dependence of the lattice constants \(a\) and \(c\) as the antimony concentration (\(x\)) increases in the unit cell. A clear trend emerges, where both lattice constants decrease linearly with increasing \(x\), indicating a direct relationship between them. This result is consistent with the findings of Yan \textit{et al.}~\cite{yan2019evolution}, where a similar behavior was observed for Mn(Bi$_{1-x}$Sb$_{x}$)$_{2}$Te$_{4}$.

\begin{figure*}
\includegraphics[width=1\textwidth]{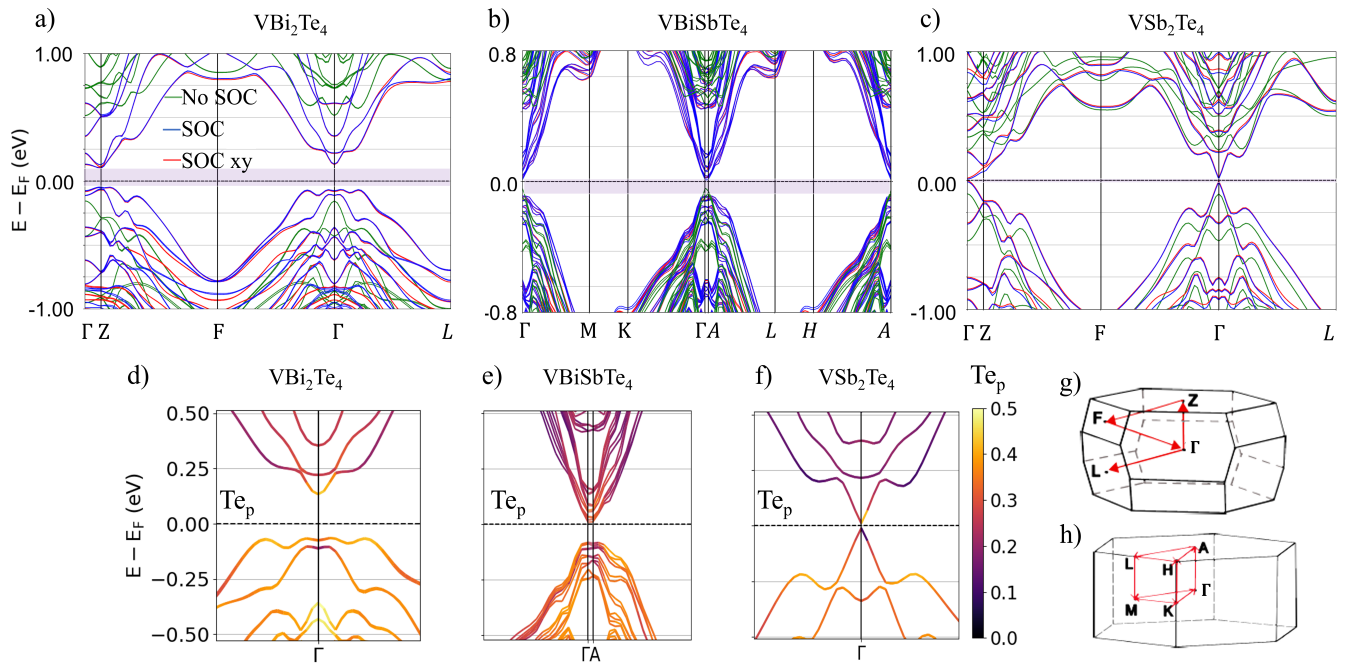}
\caption{\label{fig:bands} Band structure of V(Bi$_{1-x}$Sb$_{x}$)$_{2}$Te$_{4}$ for $x$=0, $x$=0.5, and $x$=1 in the most stable AFM configuration, with the Fermi level set as zero energy. In panels a), b), and c), the green lines represent the band structure without spin-orbit coupling (No SOC). The blue and red lines denote the band structures for out-of-plane AFM (SOC) and in-plane AFM (SOC$_{xy}$), respectively, demonstrating negligible differences near the Fermi level. (a) Band structure for VBi$_{2}$Te$_{4}$, (b) for VBiSbTe$_{4}$, and (c) for VSb$_{2}$Te$_{4}$. Panels d), e), and f) show the projected Te-$p$ orbitals in each material, highlighting the band inversion in VBi$_{2}$Te$_{4}$, VBiSbTe$_{4}$, and VSb$_{2}$Te$_{4}$, respectively. Panels g) and h) depict the first Brillouin zone paths for space groups \#166 and \#156, respectively.}
\end{figure*}


With the most stable structural phases determined, the electronic properties of VBi$_{2}$Te$_{4}$ were analyzed through band structure calculations, both with and without SOC. The bands are plotted along the $k$-paths of the rhombohedral unit cell, as shown in Figure~\ref{fig:bands}g. In both cases, a band gap separating the valence and conduction bands is observed, confirming the material’s insulating beha\-vior. In the absence of SOC, the material exhibits a direct band gap of 0.667 eV at the $\Gamma$ point. However, when SOC is included, the gap is reduced to 0.145 eV, and the band gap becomes indirect, as shown in Figure~\ref{fig:bands}a.
We also observe that the Bi-$p$ orbitals are the closest to the Fermi level and primarily contribute to the conduction band. In contrast, the Te-$p$ orbitals are more prominent near the Fermi level, with these orbitals mainly located in the valence band. Upon the activation of SOC, these orbitals undergo band inversion.

Figure~\ref{fig:bands}b shows the band structure of VBiSbTe$_{4}$, computed without and with SOC, plotted along the $k$-paths of the hexagonal Brillouin zone illustrated in Figure~\ref{fig:bands}h.  Without SOC, the material behaves as an insulator with a direct gap of 0.471 eV at the $\Gamma$ point, as shown in \mbox{Figure}~\ref{fig:bands}b. When SOC is included, the band gap decreases to 0.098 eV and becomes indirect. In both cases, the material exhibits semiconductor behavior.
A close ins\-pection of the atomic orbital contributions reveals that both bismuth and antimony atoms exhibit similar beha\-vior near the Fermi level, with their $p$-orbitals located in the conduction band. In contrast, the Te-$p$ orbitals are more concentrated in the valence band near the Fermi level. This behavior is consistent with the known band inversion in topological insulators, which typically involves the $p$-orbitals of bismuth, antimony, and tellurium atoms~\cite{zhang2009topological}.

For the study of the electronic properties of VSb$_{2}$Te$_{4}$, the band structures with and without SOC are shown in Figure~\ref{fig:bands}c. In both cases, a band gap between the valence and conduction bands is observed. In the absence of SOC, the material exhibits a direct gap of 0.338 eV at the $\Gamma$ point. However, when SOC is included, the gap is reduced to 0.021 eV, still at the $\Gamma$ point. In this case, Sb-$p$ orbitals are primarily located in the conduction band, while Te-$p$ orbitals are concentrated in the valence band. These orbitals, particularly those of bismuth, antimony, and tellurium, are responsible for the band inversion observed in materials such as Bi$_{2}$Te$_{3}$ and Sb$_{2}$Te$_{3}$\cite{zhang2009topological,zhang2010first,mazumder2021brief}.

Overall, Figure~\ref{lattices}b illustrates the variation of the band gap with and without SOC as a function of Sb concentration for the  V(Bi$_{1-x}$Sb$_{x}$)$_{2}$Te$_{4}$ materials. A clear relationship is observed between the gap and the increase in the Sb concentration, with the band gap decreasing as the value of $x$ increases. This result is consistent with similar studies where Mn occupies the V sites, such as in Ref \cite{eremeev2017competing}. 
Additionally, the SOC effect leads to a reduction of approximately 80\% in the band gap compared to the non-SOC case, across all Sb concentrations.

\begin{figure*}
\includegraphics[width=0.98\textwidth]{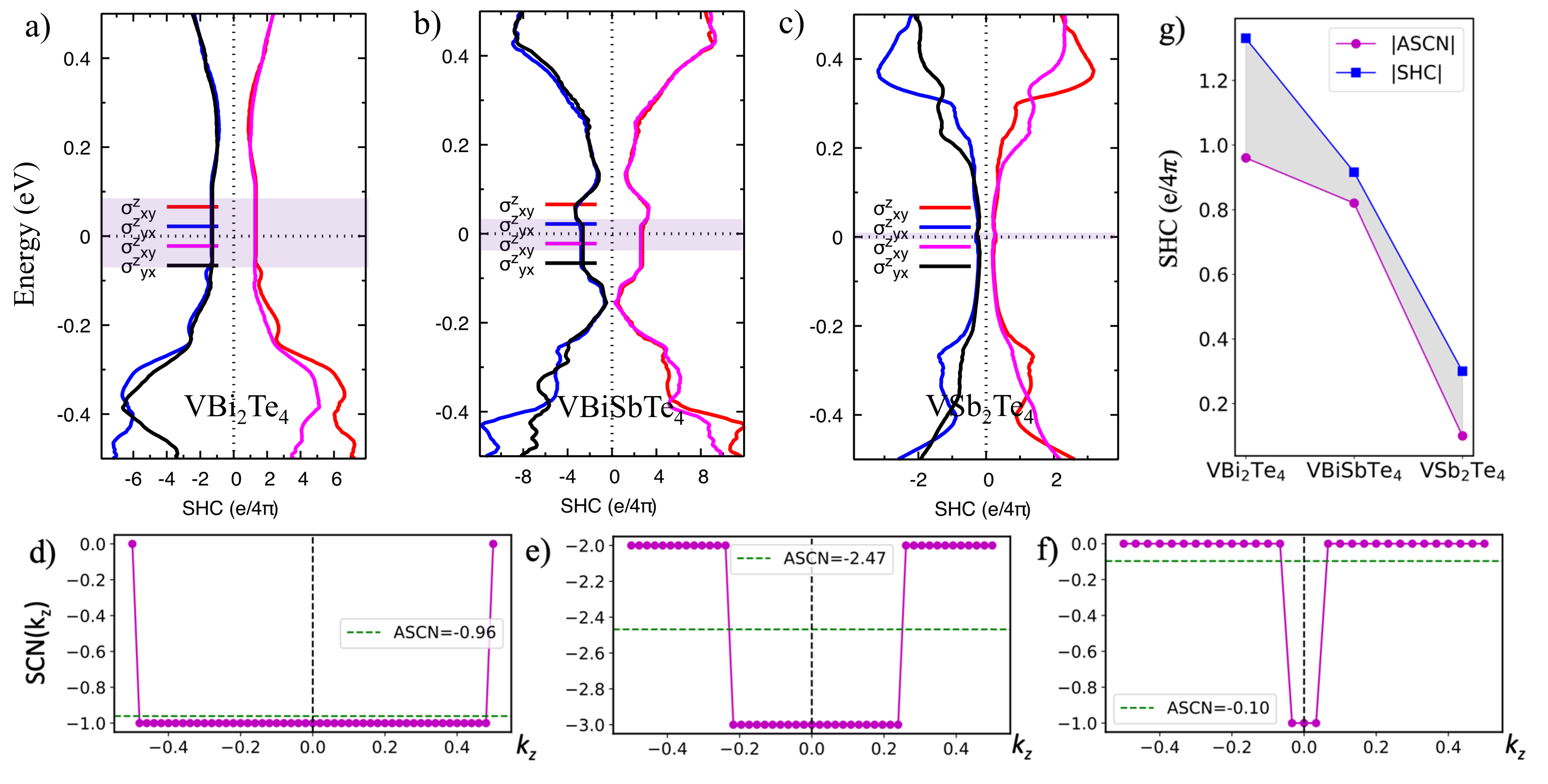}
\caption{\label{fig:topological}Spin Hall conductivity, $\sigma^{z}_{xy}$ (SHC) for the V(Bi$_{1-x}$Sb$_{x}$)$_{2}$Te$_{4}$ materials with a) $x$=0,  b) $x$=0.5 and c) $x$=1. Red and blue lines denote the SHC for out-of-plane AFM, while black and magenta correspond to in-plane AFM magnetization. SHC signal remains preserved within the band gap, independent of the Nèel vector orientation. In the shaded region, a non-zero SHC signal within the bulk band gap is found. The spin Chern number (SCN) as a function of the $k_z$ plane is shown for the V(Bi$_{1-x}$Sb$_{x}$)$_{2}$Te$_{4}$ materials with d) $x$=0,  e) $x$=0.5 and f) $x$=1; from these values, the average SCN (ASCN) can be calculated. A change in SCN by a value of 1 indicates a strong topological insulator character of this AFM materials. g) Variation of SHC and ASCN with Sb concentration in V(Bi$_{1-x}$Sb$_{x}$)$_{2}$Te$_{4}$  materials. The ASCN and SHC for VBiSbTe$_4$ were normalized per unit formula, accounting for the fact that its unit cell contains three times the number of atoms compared to VBi$_2$Te$_4$ and VSb$_2$Te$_4$. It is observed that both SHC and ASCN decrease with increasing Sb concentration.}
\end{figure*}

Regarding the magnetic properties of VBi$_{2}$Te$_{4}$, each V atom exhibits a magnetization of $2.74$ $\mu_B$, leading to an AFM2  configuration. The magnetic space group is \#167.108, the same as that of MnBi$_{2}$Te$_{4}$ material \cite{ding2020crystal}, and it exhibits symmetries such as inversion ($\mathbb{P}$) and partial-translation combined with time-reversal symmetry ($t_{1/2}\mathbb{T}$) \cite{FIND}.
For VBiSbTe$_{4}$, each V atom has a magnetization of $2.74$ $\mu_B$ along the $c$-axis, resul\-ting in an antiferromagnetic configuration (type BB). The bismuth, antimony, and tellurium atoms acquire small magnetizations that cancel out, leaving the crystal overall antiferromagnetic. This material belongs to magnetic space group \#158.60, a subgroup of P3m1, which exhibits collinear AFM with the $t_{1/2}\mathbb{T}$ symmetry but lacks inversion symmetry.
In the case of VSb$_{2}$Te$_{4}$, after structural optimization, each V atom exhibits $2.74$ $\mu_B$ with alternating magnetization along the $c$-axis. The antimony and tellurium atoms also acquire small magnetizations, resulting in an overall AFM2 configuration. This material belongs to magnetic space group \#167.108, exhibiting both  $t_{1/2}\mathbb{T}$ and $\mathbb{P}$ symmetry.

The magnetic anisotropy of V(Bi$_{1-x}$Sb$_{x}$)$_{2}$Te$_{4}$ was analyzed, revealing small energy differences between in-plane and out-of-plane AFM configurations: $-0.18$ $m$eV for VBi$_2$Te$_4$, $-0.15$ $m$eV for VBiSbTe$_4$, and $-0.16$ $m$eV for VSb$_2$Te$_4$. These results indicate a slight energetic preference for the in-plane easy axis, consistent with Petrov et al.~\cite{petrov2021domain}. However, the out-of-plane AFM magnetization configuration was chosen for analysis, as the electronic and topological properties remain unaffected by the AFM orientation. Figures~\ref{fig:bands}a–c present the SOC band structures for both in-plane (red) and out-of-plane (blue) AFM magnetization, demonstrating negligible differences near the Fermi level.

For VBi$_{2}$Te$_{4}$, Figure \ref{fig:bands}d illustrates the projection of the Te-$p$ states near the Fermi level at the $\Gamma$ point, where band inversion occurs. This inversion is indicated by the color change in the character of these orbitals, which suggests the presence of non-trivial topology in the crystal.  
By analyzing the inversion symmetry operator at the TRIMs, we find a large number of Kramers pairs with odd inversion. 
Specifically, there are three k-points with 33 Kramers pairs and one with 34 inversion pairs, which confirms the occurrence of band inversion and the material's non-trivial topological properties. The topological invariant $\mathbb{Z}_2$ is calculated to be 1, classifying the material as a strong topological insulator, in agreement with Petrov \textit{et al.} \cite{petrov2021domain}.

The topological nature of VBi$_2$Te$_4$ is corroborated by our SHC calculations, which reveal a non-zero signal within the band gap, as shown in Fig. \ref{fig:topological}a. Furthermore, Fig. \ref{fig:topological}d highlights the variation of the spin Chern number (SCN) across different k$_z$ planes. This indicator is derived from spin Wilson loop calculations on the $k_z$=$0$ and $k_z$=$\pi$ planes, yielding values of 1 and 0, respectively. The change in SCN from 0 to 1 along $k_z$ reflects the presence of spin-Weyl points and corresponds to a nontrivial $\mathbb{Z}_2$ index. The prediction of spin-Weyl points has been recently reported in the AFM material MnBi$_2$Te$_4$ ~\cite{spinweyl}. Moreover, the average SCN across all $k_z$ planes (ASCN) quantifies the topological contribution to the SHC, providing insights that are not captured by the SHC value alone.

\begin{figure*}[t]
\includegraphics[width=1.0\textwidth]{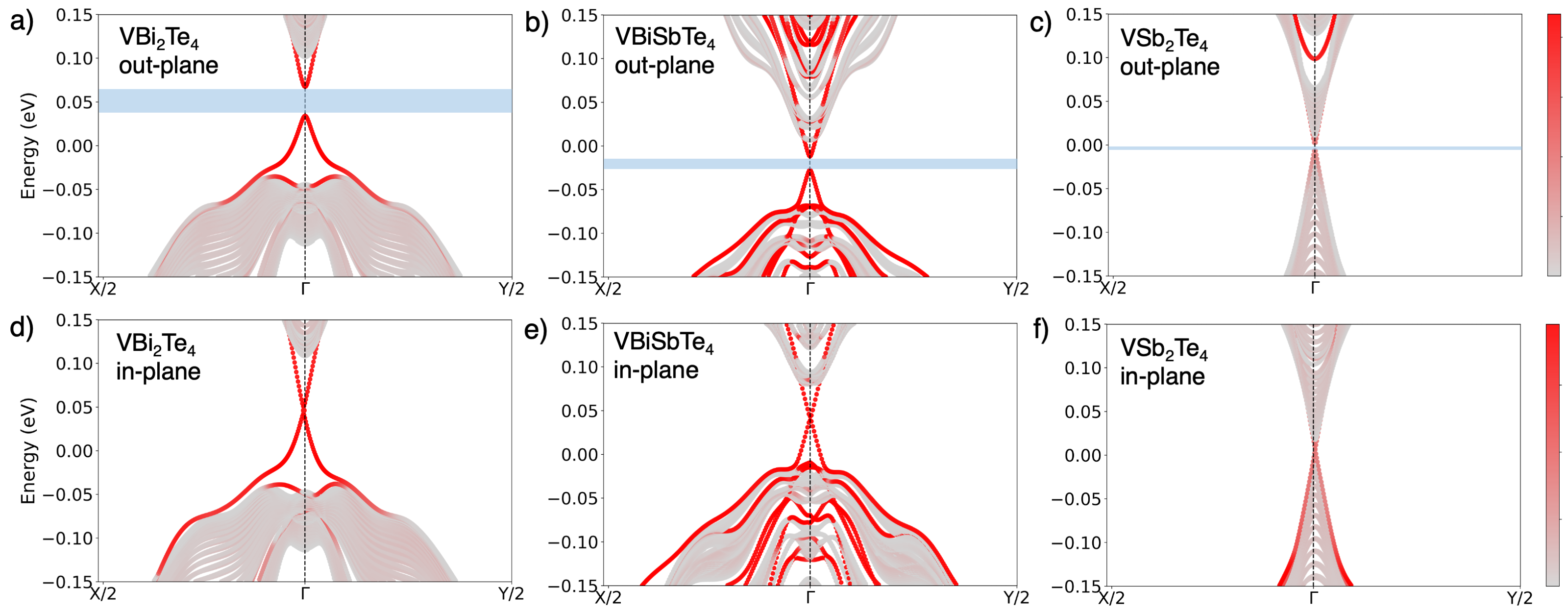}
\caption{\label{fig:surfaces}{Surface band structures projected onto the (0001) surface for V(Bi$_{1-x}$Sb$_{x}$)$_{2}$Te$_{4}$  with $x = 0$ (a,d), $x = 0.5$ (b,e), and $x = 1$ (c,f), for out-of-plane (a–c) and in-plane (d–f) magnetization configurations. Bulk-projected states are shown in gray, and topological surface states are highlighted in red. For the out-of-plane orientation, a surface band gap is observed across all compositions, consistent with the breaking of the $t_{1/2}\mathbb{T}$ symmetry on the (0001) surface. 
In contrast, in-plane magnetization yields gapless surface states with the Dirac point slightly displaced from the $\Gamma$ point in momentum space}
}
\end{figure*}

For the VBiSbTe$_{4}$ crystal, Figure \ref{fig:bands}e shows the beha\-vior of the Te-$p$ orbitals on the band structure. With the inclusion of the relativistic SOC correction, a band inversion is observed, where the states of these orbitals now occupy the conduction band, whereas previously they did not. This band inversion, as depicted in Fig. \ref{fig:bands}e, is a key indicator of non-trivial topological behavior in the material. This is further confirmed by the analysis of the topological surface states shown in Figure \ref{fig:surfaces}.

The topological behavior of VBiSbTe$_4$ is established through our SHC calculations, which reveal an SHC signal within the band gap, as shown in Fig. \ref{fig:topological}b. 
Additionally, the SCN is observed to transition from 2 on the $k_z = 0$ plane to 3 on the $k_z = \pi$ plane, as depicted in Fig. \ref{fig:topological}e. This transition in SCN serves as a definitive indicator of the material's robust topological characteris\-tics, classifying it as a strong topological insulator. The SCN variation directly reflects the nontrivial topology inherent to the VBiSbTe$_4$ compound. 


The VSb$_2$Te$_{4}$ system exhibits topological properties that are evidenced through the electronic behavior of its Te-$p$ orbitals. As shown in Figure \ref{fig:bands}f, these orbitals predominantly reside in the valence band in the absence of SOC, but upon inclusion of SOC, they shift to the conduction band at the high-symmetry $\Gamma$ point. This band inversion is a hallmark of non-trivial topology, signaling the potential for topologically protected states. To further corroborate this, the number of inversion-odd Kramers pairs at TRIMs reveals a band inversion, with three TRIMs hosting 23 odd Kramers pairs, and one TRIM showing 24 pairs. These findings yield a topological invariant $\mathbb{Z}_{2}$= 1, confirming that VSb$_2$Te$_{4}$ exhibits non-trivial topology and categorizing it as a strong magnetic topological insulator, as was described in Ref\cite{petrov2021domain}.
For this case, the topological response is confirmed by the pre\-sence of a non-zero SHC signal within the band gap, as shown in Fig. \ref{fig:topological}c. In addition, Fig. \ref{fig:topological}f illustrates the transition of the SCN from 0 on the $k_z$=$0$ plane to 1 on the $k_z$=$\pi$ plane. This change reflects the presence of spin Weyl points and is indicative of a nontrivial $\mathbb{Z}_2$ topology for the VSb$_2$Te$_{4}$ material.

The tuning of the SHC within the band gap as a function of Sb concentration in the antiferromagnetic V(Bi$_{1-x}$Sb$_{x}$)$_{2}$Te$_{4}$ materials is shown in Fig. \ref{fig:topological}g. A clear decrease in the SHC is observed as the Sb concentration increases. Furthermore, the topological contribution to the SHC, measured by the ASCN, is also presented as a function of the Sb concentration. 
The ASCN is consistently smaller than the SHC and similarly decreases with increasing Sb concentration. This tunability of spin-charge generation through the SHC via Sb concentration highlights a significant opportunity for controlling topological antiferromagnetic properties in V(Bi$_{1-x}$Sb$_{x}$)$_{2}$Te$_{4}$, which is crucial for potential device applications. Fig. \ref{fig:topological}a-c demonstrates that the SHC within the energy gap is preserved regardless of the AFM magnetization direction. These results highlight that the key electronic and topological features are robust against changes in the Nèel vector orientation.

In Figure~\ref{fig:surfaces}, we present the surface-state band structures on the (0001) surface of VBi$_{2}$Te$_{4}$, VBiSbTe$_{4}$, and VSb$_{2}$Te$_{4}$, considering both out-of-plane (a–c) and in-plane (d–f) magnetic moment orientations. In all panels, the bulk-projected states are shown in gray, while the topological surface states are highlighted in red.
For the out-of-plane configurations (Figures~\ref{fig:surfaces}a–c), a finite energy gap appears at the $\Gamma$ point, separating the valence and conduction bands. This behavior is expected, as the combined symmetry $t_{1/2}\mathbb{T}$ is broken on the (0001) surface under perpendicular magnetization. These results are consistent with previous reports on MnBi$_2$Te$_4$ exhibiting similar symmetry breaking and gap opening \cite{otrokov2019prediction,li2019magnetically,hao2019gapless,chen2019surface}. Our findings are further supported by a change in the spin Chern number, $\delta C_s = 1$, and the calculated nontrivial $Z_2$ index, confirming the presence of a topological phase.
In contrast, for the in-plane magnetic orientation (Figures~\ref{fig:surfaces}d–f), the surface states remain gapless, and the Dirac point is slightly displaced from the $\Gamma$ point in momentum space. This behavior is consistent with previous studies on magnetic topological insulators with in-plane anisotropy, where the preservation of certain symmetries—particularly mirror symmetry— prevents the opening of a surface gap \cite{petrov2021domain,li2019magnetically,hao2019gapless,chen2019surface}.

\begin{figure}
\includegraphics[width=0.485\textwidth]{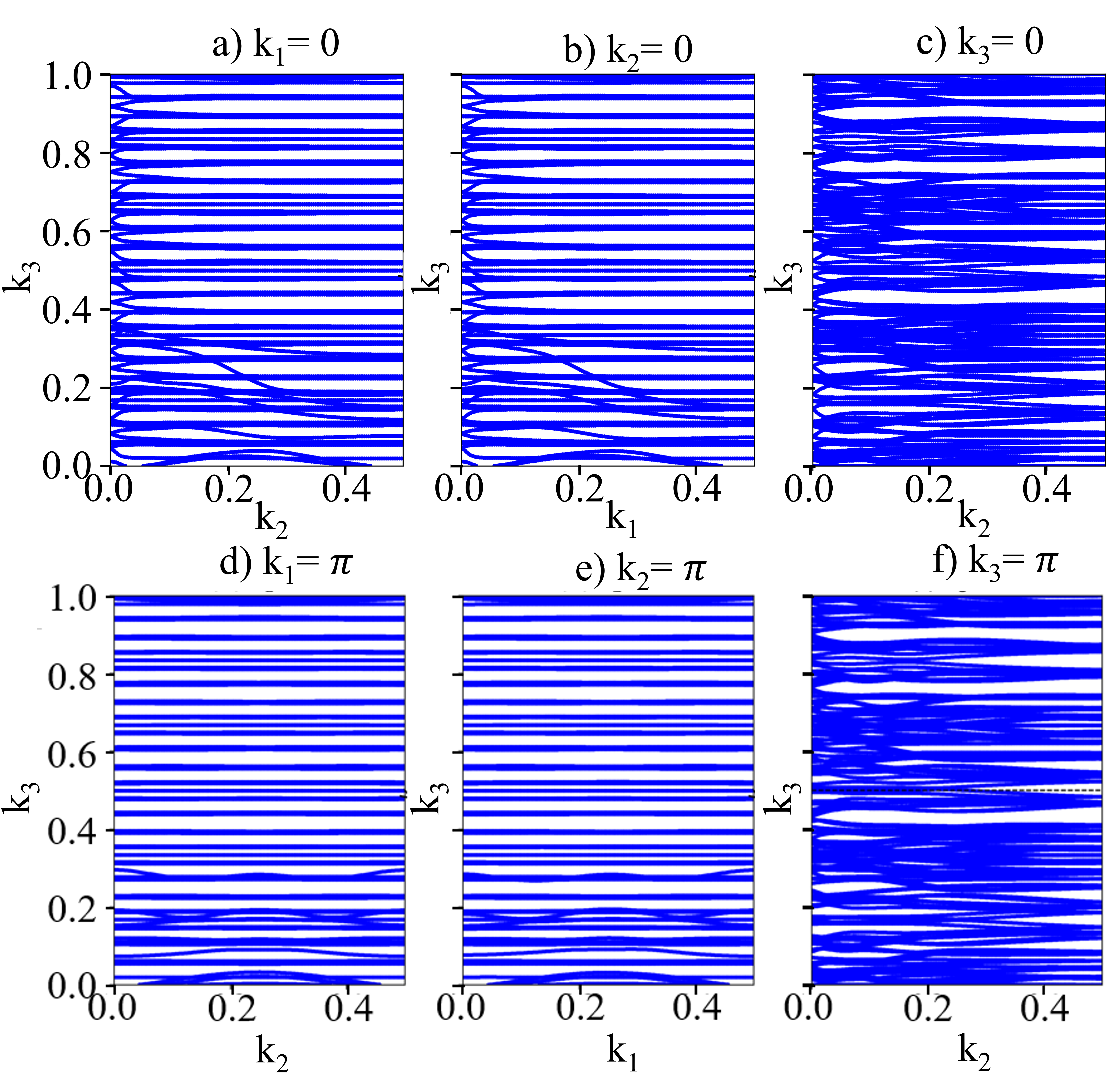}
\caption{\label{fig:wilson} Evolution of Wilson loops on six time-reversal-invariant momentum planes and corresponding $\mathbb{Z}_2$ topological indices for VBiSbTe$_4$. (a) $k_1 = 0$ plane ($k_2$-$k_3$), $\mathbb{Z}_2 = 1$; (b) $k_2 = 0$ plane ($k_1$-$k_3$), $\mathbb{Z}_2 = 1$; (c) $k_3 = 0$ plane ($k_1$-$k_3$), $\mathbb{Z}_2 = 1$; (d) $k_1 = \pi$ plane ($k_2$-$k_3$), $\mathbb{Z}_2 = 0$; (e) $k_2 = \pi$ plane ($k_1$-$k_3$), $\mathbb{Z}_2 = 0$; (f) $k_3 = \pi$ plane ($k_1$-$k_3$), $\mathbb{Z}_2 = 0$. The Wilson loop spectrum crosses the $k_i$ horizontal line an odd number of times in (a), (b), and (c), leading to $\mathbb{Z}_2 = 1$.}
\end{figure}

Finally, to confirm the nontrivial topology of the low-energy electronic states in VBiSbTe$_4$, we computed the $\mathbb{Z}_2$ topological invariants using the Wilson loop (Wannier charge center) method, as implemented in \texttt{WannierTools}\cite{WU2017}. The calculations were performed for the six time-reversal-invariant momentum (TRIM) planes, shown in Fig.\ref{fig:wilson}: (a) $k_1 = 0.0$, (b) $k_2 = 0.0$, (c) $k_3 = 0.0$, (d) $k_1 = 0.5$, (e) $k_2 = 0.5$, and (f) $k_3 = 0.5$, where $k_1$, $k_2$, and $k_3$ are expressed in units of the reciprocal lattice vectors. 
It is well established that the $\mathbb{Z}_2$ invariant is determined by the parity of Wilson loop crossings relative to a reference horizontal line: an odd number of crossings corresponds to $\mathbb{Z}_2 = 1$, while an even number results in $\mathbb{Z}_2 = 0$. Based on this criterion, our calculations yield $\mathbb{Z}_2 = 1$ for the $k_1 = 0.0$, $k_2 = 0.0$, and $k_3 = 0.0$ planes, and $\mathbb{Z}_2 = 0$ for the $k_1 = 0.5$, $k_2 = 0.5$, and $k_3 = 0.5$ planes. Consequently, the overall topological indices of the system are given by $v_0; (v_1 v_2 v_3) = 1;(000)$, confirming that VBiSbTe$_4$ is a strong topological material.

\section{\label{sec:level222}Conclusions}

In this study, we investigated the structural, electronic, magnetic, and topological properties of the la\-yered topological insulators V(Bi$_{1-x}$Sb$_{x}$)$_{2}$Te$_{4}$ for compositions $x$=0, $x$=0.5, and $x$=1. The stoichiometric formula enabled us to explore various configurations of bismuth and antimony layers, each associated with distinct magnetic arrangements depending on the Sb concentration. Using $ab$-$initio$ calculations, we determined that the most energetically stable configuration corresponds to an antiferromagnetic arrangement for all compositions. 
Speci\-fically, for $x$=$0$ and $x$=$1$, the AFM2 configurations consistently yielded the lowest energy, in agreement with previous studies. In all cases, we found that the materials exhibit semiconducting behavior, with a band gap that is significantly reduced by inclusion of spin-orbit coupling.

Topological analysis revealed band inversion, a characteristic feature of topologically non-trivial phases, and the topological invariant $\mathbb{Z}_2$ confirmed that the materials exhibit strong topological behavior.
Our findings cla\-ssify the studied materials as strong topological insulators, similar to other well-known topological phases, such as those in MnBi$_{2}$Te$_{4}$ and Bi$_{2}$Te$_{3}$.
The tunability of the spin Hall conductivity and the average spin Chern number with Sb concentration present a promising approach to control spin-charge generation. 
The topological surface states observed across Sb compositions, coupled with the bulk protection provided by partial-translation and time-reversal symmetries, confirm the topological nature of these materials. Our findings establish V(Bi$_{1-x}$Sb$_{x}$)$_{2}$Te$_{4}$ as promising candidates for next-generation AFM topological devices.

\begin{acknowledgments}
The authors wish to acknowledge the Physics of Novel Materials Group and the DIEB of the Universidad Nacional de Colombia, within the project code Hermes 60672. The calculations reported in this work were performed using GRANADO and HPC facilities at the Universidad del Norte.
\end{acknowledgments}

\nocite{*}

\bibliography{apssamp}

\begin{thebibliography}{57}%
\makeatletter
\providecommand \@ifxundefined [1]{%
 \@ifx{#1\undefined}
}%
\providecommand \@ifnum [1]{%
 \ifnum #1\expandafter \@firstoftwo
 \else \expandafter \@secondoftwo
 \fi
}%
\providecommand \@ifx [1]{%
 \ifx #1\expandafter \@firstoftwo
 \else \expandafter \@secondoftwo
 \fi
}%
\providecommand \natexlab [1]{#1}%
\providecommand \enquote  [1]{``#1''}%
\providecommand \bibnamefont  [1]{#1}%
\providecommand \bibfnamefont [1]{#1}%
\providecommand \citenamefont [1]{#1}%
\providecommand \href@noop [0]{\@secondoftwo}%
\providecommand \href [0]{\begingroup \@sanitize@url \@href}%
\providecommand \@href[1]{\@@startlink{#1}\@@href}%
\providecommand \@@href[1]{\endgroup#1\@@endlink}%
\providecommand \@sanitize@url [0]{\catcode `\\12\catcode `\$12\catcode
  `\&12\catcode `\#12\catcode `\^12\catcode `\_12\catcode `\%12\relax}%
\providecommand \@@startlink[1]{}%
\providecommand \@@endlink[0]{}%
\providecommand \url  [0]{\begingroup\@sanitize@url \@url }%
\providecommand \@url [1]{\endgroup\@href {#1}{\urlprefix }}%
\providecommand \urlprefix  [0]{URL }%
\providecommand \Eprint [0]{\href }%
\providecommand \doibase [0]{https://doi.org/}%
\providecommand \selectlanguage [0]{\@gobble}%
\providecommand \bibinfo  [0]{\@secondoftwo}%
\providecommand \bibfield  [0]{\@secondoftwo}%
\providecommand \translation [1]{[#1]}%
\providecommand \BibitemOpen [0]{}%
\providecommand \bibitemStop [0]{}%
\providecommand \bibitemNoStop [0]{.\EOS\space}%
\providecommand \EOS [0]{\spacefactor3000\relax}%
\providecommand \BibitemShut  [1]{\csname bibitem#1\endcsname}%
\let\auto@bib@innerbib\@empty
\bibitem [{\citenamefont {Qi}\ and\ \citenamefont
  {Zhang}(2011)}]{qi2011topological}%
  \BibitemOpen
  \bibfield  {author} {\bibinfo {author} {\bibfnamefont {X.-L.}\ \bibnamefont
  {Qi}}\ and\ \bibinfo {author} {\bibfnamefont {S.-C.}\ \bibnamefont {Zhang}},\
  }\bibfield  {title} {\bibinfo {title} {Topological insulators and
  superconductors},\ }\href@noop {} {\bibfield  {journal} {\bibinfo  {journal}
  {Reviews of modern physics}\ }\textbf {\bibinfo {volume} {83}},\ \bibinfo
  {pages} {1057} (\bibinfo {year} {2011})}\BibitemShut {NoStop}%
\bibitem [{\citenamefont {Kumar}\ \emph {et~al.}(2020)\citenamefont {Kumar},
  \citenamefont {Guin}, \citenamefont {Manna}, \citenamefont {Shekhar},\ and\
  \citenamefont {Felser}}]{kumar2020topological}%
  \BibitemOpen
  \bibfield  {author} {\bibinfo {author} {\bibfnamefont {N.}~\bibnamefont
  {Kumar}}, \bibinfo {author} {\bibfnamefont {S.~N.}\ \bibnamefont {Guin}},
  \bibinfo {author} {\bibfnamefont {K.}~\bibnamefont {Manna}}, \bibinfo
  {author} {\bibfnamefont {C.}~\bibnamefont {Shekhar}},\ and\ \bibinfo {author}
  {\bibfnamefont {C.}~\bibnamefont {Felser}},\ }\bibfield  {title} {\bibinfo
  {title} {Topological quantum materials from the viewpoint of chemistry},\
  }\href@noop {} {\bibfield  {journal} {\bibinfo  {journal} {Chemical Reviews}\
  }\textbf {\bibinfo {volume} {121}},\ \bibinfo {pages} {2780} (\bibinfo {year}
  {2020})}\BibitemShut {NoStop}%
\bibitem [{\citenamefont {Xiao}\ and\ \citenamefont
  {Yan}(2021)}]{xiao2021first}%
  \BibitemOpen
  \bibfield  {author} {\bibinfo {author} {\bibfnamefont {J.}~\bibnamefont
  {Xiao}}\ and\ \bibinfo {author} {\bibfnamefont {B.}~\bibnamefont {Yan}},\
  }\bibfield  {title} {\bibinfo {title} {First-principles calculations for
  topological quantum materials},\ }\href@noop {} {\bibfield  {journal}
  {\bibinfo  {journal} {Nature reviews physics}\ }\textbf {\bibinfo {volume}
  {3}},\ \bibinfo {pages} {283} (\bibinfo {year} {2021})}\BibitemShut {NoStop}%
\bibitem [{\citenamefont {Cava}\ \emph {et~al.}(2013)\citenamefont {Cava},
  \citenamefont {Ji}, \citenamefont {Fuccillo}, \citenamefont {Gibson},\ and\
  \citenamefont {Hor}}]{cava2013crystal}%
  \BibitemOpen
  \bibfield  {author} {\bibinfo {author} {\bibfnamefont {R.~J.}\ \bibnamefont
  {Cava}}, \bibinfo {author} {\bibfnamefont {H.}~\bibnamefont {Ji}}, \bibinfo
  {author} {\bibfnamefont {M.~K.}\ \bibnamefont {Fuccillo}}, \bibinfo {author}
  {\bibfnamefont {Q.~D.}\ \bibnamefont {Gibson}},\ and\ \bibinfo {author}
  {\bibfnamefont {Y.~S.}\ \bibnamefont {Hor}},\ }\bibfield  {title} {\bibinfo
  {title} {Crystal structure and chemistry of topological insulators},\
  }\href@noop {} {\bibfield  {journal} {\bibinfo  {journal} {Journal of
  Materials Chemistry C}\ }\textbf {\bibinfo {volume} {1}},\ \bibinfo {pages}
  {3176} (\bibinfo {year} {2013})}\BibitemShut {NoStop}%
\bibitem [{\citenamefont {Yu}\ \emph {et~al.}(2010)\citenamefont {Yu},
  \citenamefont {Zhang}, \citenamefont {Zhang}, \citenamefont {Zhang},
  \citenamefont {Dai},\ and\ \citenamefont {Fang}}]{yu2010quantized}%
  \BibitemOpen
  \bibfield  {author} {\bibinfo {author} {\bibfnamefont {R.}~\bibnamefont
  {Yu}}, \bibinfo {author} {\bibfnamefont {W.}~\bibnamefont {Zhang}}, \bibinfo
  {author} {\bibfnamefont {H.-J.}\ \bibnamefont {Zhang}}, \bibinfo {author}
  {\bibfnamefont {S.-C.}\ \bibnamefont {Zhang}}, \bibinfo {author}
  {\bibfnamefont {X.}~\bibnamefont {Dai}},\ and\ \bibinfo {author}
  {\bibfnamefont {Z.}~\bibnamefont {Fang}},\ }\bibfield  {title} {\bibinfo
  {title} {Quantized anomalous hall effect in magnetic topological
  insulators},\ }\href@noop {} {\bibfield  {journal} {\bibinfo  {journal}
  {science}\ }\textbf {\bibinfo {volume} {329}},\ \bibinfo {pages} {61}
  (\bibinfo {year} {2010})}\BibitemShut {NoStop}%
\bibitem [{\citenamefont {Zou}\ \emph {et~al.}(2017)\citenamefont {Zou},
  \citenamefont {Wang}, \citenamefont {Kou}, \citenamefont {Lang},
  \citenamefont {Fan}, \citenamefont {Choi}, \citenamefont {Fedorov},
  \citenamefont {Wang}, \citenamefont {He}, \citenamefont {Xu} \emph
  {et~al.}}]{zou2017observation}%
  \BibitemOpen
  \bibfield  {author} {\bibinfo {author} {\bibfnamefont {W.}~\bibnamefont
  {Zou}}, \bibinfo {author} {\bibfnamefont {W.}~\bibnamefont {Wang}}, \bibinfo
  {author} {\bibfnamefont {X.}~\bibnamefont {Kou}}, \bibinfo {author}
  {\bibfnamefont {M.}~\bibnamefont {Lang}}, \bibinfo {author} {\bibfnamefont
  {Y.}~\bibnamefont {Fan}}, \bibinfo {author} {\bibfnamefont {E.~S.}\
  \bibnamefont {Choi}}, \bibinfo {author} {\bibfnamefont {A.~V.}\ \bibnamefont
  {Fedorov}}, \bibinfo {author} {\bibfnamefont {K.}~\bibnamefont {Wang}},
  \bibinfo {author} {\bibfnamefont {L.}~\bibnamefont {He}}, \bibinfo {author}
  {\bibfnamefont {Y.}~\bibnamefont {Xu}}, \emph {et~al.},\ }\bibfield  {title}
  {\bibinfo {title} {Observation of quantum hall effect in an ultra-thin (bi0.
  53sb0. 47) 2te3 film},\ }\href@noop {} {\bibfield  {journal} {\bibinfo
  {journal} {Applied Physics Letters}\ }\textbf {\bibinfo {volume} {110}}
  (\bibinfo {year} {2017})}\BibitemShut {NoStop}%
\bibitem [{\citenamefont {Konig}\ \emph {et~al.}(2007)\citenamefont {Konig},
  \citenamefont {Wiedmann}, \citenamefont {Brune}, \citenamefont {Roth},
  \citenamefont {Buhmann}, \citenamefont {Molenkamp}, \citenamefont {Qi},\ and\
  \citenamefont {Zhang}}]{konig2007quantum}%
  \BibitemOpen
  \bibfield  {author} {\bibinfo {author} {\bibfnamefont {M.}~\bibnamefont
  {Konig}}, \bibinfo {author} {\bibfnamefont {S.}~\bibnamefont {Wiedmann}},
  \bibinfo {author} {\bibfnamefont {C.}~\bibnamefont {Brune}}, \bibinfo
  {author} {\bibfnamefont {A.}~\bibnamefont {Roth}}, \bibinfo {author}
  {\bibfnamefont {H.}~\bibnamefont {Buhmann}}, \bibinfo {author} {\bibfnamefont
  {L.~W.}\ \bibnamefont {Molenkamp}}, \bibinfo {author} {\bibfnamefont {X.-L.}\
  \bibnamefont {Qi}},\ and\ \bibinfo {author} {\bibfnamefont {S.-C.}\
  \bibnamefont {Zhang}},\ }\bibfield  {title} {\bibinfo {title} {Quantum spin
  hall insulator state in hgte quantum wells},\ }\href@noop {} {\bibfield
  {journal} {\bibinfo  {journal} {Science}\ }\textbf {\bibinfo {volume}
  {318}},\ \bibinfo {pages} {766} (\bibinfo {year} {2007})}\BibitemShut
  {NoStop}%
\bibitem [{\citenamefont {Yan}\ and\ \citenamefont
  {Felser}(2017)}]{yan2017topological}%
  \BibitemOpen
  \bibfield  {author} {\bibinfo {author} {\bibfnamefont {B.}~\bibnamefont
  {Yan}}\ and\ \bibinfo {author} {\bibfnamefont {C.}~\bibnamefont {Felser}},\
  }\bibfield  {title} {\bibinfo {title} {Topological materials: Weyl
  semimetals},\ }\href@noop {} {\bibfield  {journal} {\bibinfo  {journal}
  {Annual Review of Condensed Matter Physics}\ }\textbf {\bibinfo {volume}
  {8}},\ \bibinfo {pages} {337} (\bibinfo {year} {2017})}\BibitemShut {NoStop}%
\bibitem [{\citenamefont {Zhang}\ \emph {et~al.}(2009)\citenamefont {Zhang},
  \citenamefont {Liu}, \citenamefont {Qi}, \citenamefont {Dai}, \citenamefont
  {Fang},\ and\ \citenamefont {Zhang}}]{zhang2009topological}%
  \BibitemOpen
  \bibfield  {author} {\bibinfo {author} {\bibfnamefont {H.}~\bibnamefont
  {Zhang}}, \bibinfo {author} {\bibfnamefont {C.-X.}\ \bibnamefont {Liu}},
  \bibinfo {author} {\bibfnamefont {X.-L.}\ \bibnamefont {Qi}}, \bibinfo
  {author} {\bibfnamefont {X.}~\bibnamefont {Dai}}, \bibinfo {author}
  {\bibfnamefont {Z.}~\bibnamefont {Fang}},\ and\ \bibinfo {author}
  {\bibfnamefont {S.-C.}\ \bibnamefont {Zhang}},\ }\bibfield  {title} {\bibinfo
  {title} {Topological insulators in bi2se3, bi2te3 and sb2te3 with a single
  dirac cone on the surface},\ }\href@noop {} {\bibfield  {journal} {\bibinfo
  {journal} {Nature physics}\ }\textbf {\bibinfo {volume} {5}},\ \bibinfo
  {pages} {438} (\bibinfo {year} {2009})}\BibitemShut {NoStop}%
\bibitem [{\citenamefont {Zhang}\ \emph {et~al.}(2010)\citenamefont {Zhang},
  \citenamefont {Yu}, \citenamefont {Zhang}, \citenamefont {Dai},\ and\
  \citenamefont {Fang}}]{zhang2010first}%
  \BibitemOpen
  \bibfield  {author} {\bibinfo {author} {\bibfnamefont {W.}~\bibnamefont
  {Zhang}}, \bibinfo {author} {\bibfnamefont {R.}~\bibnamefont {Yu}}, \bibinfo
  {author} {\bibfnamefont {H.-J.}\ \bibnamefont {Zhang}}, \bibinfo {author}
  {\bibfnamefont {X.}~\bibnamefont {Dai}},\ and\ \bibinfo {author}
  {\bibfnamefont {Z.}~\bibnamefont {Fang}},\ }\bibfield  {title} {\bibinfo
  {title} {First-principles studies of the three-dimensional strong topological
  insulators bi2te3, bi2se3 and sb2te3},\ }\href@noop {} {\bibfield  {journal}
  {\bibinfo  {journal} {New Journal of Physics}\ }\textbf {\bibinfo {volume}
  {12}},\ \bibinfo {pages} {065013} (\bibinfo {year} {2010})}\BibitemShut
  {NoStop}%
\bibitem [{\citenamefont {Mazumder}\ and\ \citenamefont
  {Shirage}(2021)}]{mazumder2021brief}%
  \BibitemOpen
  \bibfield  {author} {\bibinfo {author} {\bibfnamefont {K.}~\bibnamefont
  {Mazumder}}\ and\ \bibinfo {author} {\bibfnamefont {P.~M.}\ \bibnamefont
  {Shirage}},\ }\bibfield  {title} {\bibinfo {title} {A brief review of bi2se3
  based topological insulator: From fundamentals to applications},\ }\href@noop
  {} {\bibfield  {journal} {\bibinfo  {journal} {Journal of Alloys and
  Compounds}\ }\textbf {\bibinfo {volume} {888}},\ \bibinfo {pages} {161492}
  (\bibinfo {year} {2021})}\BibitemShut {NoStop}%
\bibitem [{\citenamefont {He}\ \emph {et~al.}(2013)\citenamefont {He},
  \citenamefont {Kou},\ and\ \citenamefont {Wang}}]{he2013review}%
  \BibitemOpen
  \bibfield  {author} {\bibinfo {author} {\bibfnamefont {L.}~\bibnamefont
  {He}}, \bibinfo {author} {\bibfnamefont {X.}~\bibnamefont {Kou}},\ and\
  \bibinfo {author} {\bibfnamefont {K.~L.}\ \bibnamefont {Wang}},\ }\href@noop
  {} {\bibinfo {title} {Review of 3d topological insulator thin-film growth by
  molecular beam epitaxy and potential applications}} (\bibinfo {year}
  {2013})\BibitemShut {NoStop}%
\bibitem [{\citenamefont {Chang}\ \emph {et~al.}(2015)\citenamefont {Chang},
  \citenamefont {Zhao}, \citenamefont {Kim}, \citenamefont {Zhang},
  \citenamefont {Assaf}, \citenamefont {Heiman}, \citenamefont {Zhang},
  \citenamefont {Liu}, \citenamefont {Chan},\ and\ \citenamefont
  {Moodera}}]{chang2015high}%
  \BibitemOpen
  \bibfield  {author} {\bibinfo {author} {\bibfnamefont {C.-Z.}\ \bibnamefont
  {Chang}}, \bibinfo {author} {\bibfnamefont {W.}~\bibnamefont {Zhao}},
  \bibinfo {author} {\bibfnamefont {D.~Y.}\ \bibnamefont {Kim}}, \bibinfo
  {author} {\bibfnamefont {H.}~\bibnamefont {Zhang}}, \bibinfo {author}
  {\bibfnamefont {B.~A.}\ \bibnamefont {Assaf}}, \bibinfo {author}
  {\bibfnamefont {D.}~\bibnamefont {Heiman}}, \bibinfo {author} {\bibfnamefont
  {S.-C.}\ \bibnamefont {Zhang}}, \bibinfo {author} {\bibfnamefont
  {C.}~\bibnamefont {Liu}}, \bibinfo {author} {\bibfnamefont {M.~H.}\
  \bibnamefont {Chan}},\ and\ \bibinfo {author} {\bibfnamefont {J.~S.}\
  \bibnamefont {Moodera}},\ }\bibfield  {title} {\bibinfo {title}
  {High-precision realization of robust quantum anomalous hall state in a hard
  ferromagnetic topological insulator},\ }\href@noop {} {\bibfield  {journal}
  {\bibinfo  {journal} {Nature materials}\ }\textbf {\bibinfo {volume} {14}},\
  \bibinfo {pages} {473} (\bibinfo {year} {2015})}\BibitemShut {NoStop}%
\bibitem [{\citenamefont {Otrokov}\ \emph {et~al.}(2019)\citenamefont
  {Otrokov}, \citenamefont {Klimovskikh}, \citenamefont {Bentmann},
  \citenamefont {Estyunin}, \citenamefont {Zeugner}, \citenamefont {Aliev},
  \citenamefont {Ga{\ss}}, \citenamefont {Wolter}, \citenamefont {Koroleva},
  \citenamefont {Shikin} \emph {et~al.}}]{otrokov2019prediction}%
  \BibitemOpen
  \bibfield  {author} {\bibinfo {author} {\bibfnamefont {M.~M.}\ \bibnamefont
  {Otrokov}}, \bibinfo {author} {\bibfnamefont {I.~I.}\ \bibnamefont
  {Klimovskikh}}, \bibinfo {author} {\bibfnamefont {H.}~\bibnamefont
  {Bentmann}}, \bibinfo {author} {\bibfnamefont {D.}~\bibnamefont {Estyunin}},
  \bibinfo {author} {\bibfnamefont {A.}~\bibnamefont {Zeugner}}, \bibinfo
  {author} {\bibfnamefont {Z.~S.}\ \bibnamefont {Aliev}}, \bibinfo {author}
  {\bibfnamefont {S.}~\bibnamefont {Ga{\ss}}}, \bibinfo {author} {\bibfnamefont
  {A.}~\bibnamefont {Wolter}}, \bibinfo {author} {\bibfnamefont
  {A.}~\bibnamefont {Koroleva}}, \bibinfo {author} {\bibfnamefont {A.~M.}\
  \bibnamefont {Shikin}}, \emph {et~al.},\ }\bibfield  {title} {\bibinfo
  {title} {Prediction and observation of an antiferromagnetic topological
  insulator},\ }\href@noop {} {\bibfield  {journal} {\bibinfo  {journal}
  {Nature}\ }\textbf {\bibinfo {volume} {576}},\ \bibinfo {pages} {416}
  (\bibinfo {year} {2019})}\BibitemShut {NoStop}%
\bibitem [{\citenamefont {Wu}\ \emph {et~al.}(2019)\citenamefont {Wu},
  \citenamefont {Liu}, \citenamefont {Sasase}, \citenamefont {Ienaga},
  \citenamefont {Obata}, \citenamefont {Yukawa}, \citenamefont {Horiba},
  \citenamefont {Kumigashira}, \citenamefont {Okuma}, \citenamefont {Inoshita}
  \emph {et~al.}}]{wu2019natural}%
  \BibitemOpen
  \bibfield  {author} {\bibinfo {author} {\bibfnamefont {J.}~\bibnamefont
  {Wu}}, \bibinfo {author} {\bibfnamefont {F.}~\bibnamefont {Liu}}, \bibinfo
  {author} {\bibfnamefont {M.}~\bibnamefont {Sasase}}, \bibinfo {author}
  {\bibfnamefont {K.}~\bibnamefont {Ienaga}}, \bibinfo {author} {\bibfnamefont
  {Y.}~\bibnamefont {Obata}}, \bibinfo {author} {\bibfnamefont
  {R.}~\bibnamefont {Yukawa}}, \bibinfo {author} {\bibfnamefont
  {K.}~\bibnamefont {Horiba}}, \bibinfo {author} {\bibfnamefont
  {H.}~\bibnamefont {Kumigashira}}, \bibinfo {author} {\bibfnamefont
  {S.}~\bibnamefont {Okuma}}, \bibinfo {author} {\bibfnamefont
  {T.}~\bibnamefont {Inoshita}}, \emph {et~al.},\ }\bibfield  {title} {\bibinfo
  {title} {Natural van der waals heterostructural single crystals with both
  magnetic and topological properties},\ }\href@noop {} {\bibfield  {journal}
  {\bibinfo  {journal} {Science advances}\ }\textbf {\bibinfo {volume} {5}},\
  \bibinfo {pages} {eaax9989} (\bibinfo {year} {2019})}\BibitemShut {NoStop}%
\bibitem [{\citenamefont {Eremeev}\ \emph {et~al.}(2017)\citenamefont
  {Eremeev}, \citenamefont {Otrokov},\ and\ \citenamefont
  {Chulkov}}]{eremeev2017competing}%
  \BibitemOpen
  \bibfield  {author} {\bibinfo {author} {\bibfnamefont {S.~V.}\ \bibnamefont
  {Eremeev}}, \bibinfo {author} {\bibfnamefont {M.}~\bibnamefont {Otrokov}},\
  and\ \bibinfo {author} {\bibfnamefont {E.~V.}\ \bibnamefont {Chulkov}},\
  }\bibfield  {title} {\bibinfo {title} {Competing rhombohedral and monoclinic
  crystal structures in mnpn2ch4 compounds: An ab-initio study},\ }\href@noop
  {} {\bibfield  {journal} {\bibinfo  {journal} {Journal of Alloys and
  Compounds}\ }\textbf {\bibinfo {volume} {709}},\ \bibinfo {pages} {172}
  (\bibinfo {year} {2017})}\BibitemShut {NoStop}%
\bibitem [{\citenamefont {Chen}\ \emph {et~al.}(2020)\citenamefont {Chen},
  \citenamefont {Wang}, \citenamefont {Shi}, \citenamefont {Jiang},
  \citenamefont {Liu}, \citenamefont {Cui}, \citenamefont {Zhang},\ and\
  \citenamefont {Li}}]{chen2020electronic}%
  \BibitemOpen
  \bibfield  {author} {\bibinfo {author} {\bibfnamefont {L.}~\bibnamefont
  {Chen}}, \bibinfo {author} {\bibfnamefont {D.}~\bibnamefont {Wang}}, \bibinfo
  {author} {\bibfnamefont {C.}~\bibnamefont {Shi}}, \bibinfo {author}
  {\bibfnamefont {C.}~\bibnamefont {Jiang}}, \bibinfo {author} {\bibfnamefont
  {H.}~\bibnamefont {Liu}}, \bibinfo {author} {\bibfnamefont {G.}~\bibnamefont
  {Cui}}, \bibinfo {author} {\bibfnamefont {X.}~\bibnamefont {Zhang}},\ and\
  \bibinfo {author} {\bibfnamefont {X.}~\bibnamefont {Li}},\ }\bibfield
  {title} {\bibinfo {title} {Electronic structure and magnetism of mnsb 2 te
  4},\ }\href@noop {} {\bibfield  {journal} {\bibinfo  {journal} {Journal of
  Materials Science}\ }\textbf {\bibinfo {volume} {55}},\ \bibinfo {pages}
  {14292} (\bibinfo {year} {2020})}\BibitemShut {NoStop}%
\bibitem [{\citenamefont {Zhou}\ \emph {et~al.}(2020)\citenamefont {Zhou},
  \citenamefont {Tan}, \citenamefont {Yan}, \citenamefont {Fang}, \citenamefont
  {Shi},\ and\ \citenamefont {Weng}}]{zhou2020topological}%
  \BibitemOpen
  \bibfield  {author} {\bibinfo {author} {\bibfnamefont {L.}~\bibnamefont
  {Zhou}}, \bibinfo {author} {\bibfnamefont {Z.}~\bibnamefont {Tan}}, \bibinfo
  {author} {\bibfnamefont {D.}~\bibnamefont {Yan}}, \bibinfo {author}
  {\bibfnamefont {Z.}~\bibnamefont {Fang}}, \bibinfo {author} {\bibfnamefont
  {Y.}~\bibnamefont {Shi}},\ and\ \bibinfo {author} {\bibfnamefont
  {H.}~\bibnamefont {Weng}},\ }\bibfield  {title} {\bibinfo {title}
  {Topological phase transition in the layered magnetic compound mnsb 2 te 4:
  Spin-orbit coupling and interlayer coupling dependence},\ }\href@noop {}
  {\bibfield  {journal} {\bibinfo  {journal} {Physical review B}\ }\textbf
  {\bibinfo {volume} {102}},\ \bibinfo {pages} {085114} (\bibinfo {year}
  {2020})}\BibitemShut {NoStop}%
\bibitem [{\citenamefont {Lei}\ \emph {et~al.}(2020)\citenamefont {Lei},
  \citenamefont {Chen},\ and\ \citenamefont {MacDonald}}]{lei2020magnetized}%
  \BibitemOpen
  \bibfield  {author} {\bibinfo {author} {\bibfnamefont {C.}~\bibnamefont
  {Lei}}, \bibinfo {author} {\bibfnamefont {S.}~\bibnamefont {Chen}},\ and\
  \bibinfo {author} {\bibfnamefont {A.~H.}\ \bibnamefont {MacDonald}},\
  }\bibfield  {title} {\bibinfo {title} {Magnetized topological insulator
  multilayers},\ }\href@noop {} {\bibfield  {journal} {\bibinfo  {journal}
  {Proceedings of the National Academy of Sciences}\ }\textbf {\bibinfo
  {volume} {117}},\ \bibinfo {pages} {27224} (\bibinfo {year}
  {2020})}\BibitemShut {NoStop}%
\bibitem [{\citenamefont {Lee}\ \emph {et~al.}(2021)\citenamefont {Lee},
  \citenamefont {Graf}, \citenamefont {Min}, \citenamefont {Zhu}, \citenamefont
  {Yi}, \citenamefont {Ciocys}, \citenamefont {Wang}, \citenamefont {Choi},
  \citenamefont {Basnet}, \citenamefont {Fereidouni} \emph
  {et~al.}}]{lee2021evidence}%
  \BibitemOpen
  \bibfield  {author} {\bibinfo {author} {\bibfnamefont {S.~H.}\ \bibnamefont
  {Lee}}, \bibinfo {author} {\bibfnamefont {D.}~\bibnamefont {Graf}}, \bibinfo
  {author} {\bibfnamefont {L.}~\bibnamefont {Min}}, \bibinfo {author}
  {\bibfnamefont {Y.}~\bibnamefont {Zhu}}, \bibinfo {author} {\bibfnamefont
  {H.}~\bibnamefont {Yi}}, \bibinfo {author} {\bibfnamefont {S.}~\bibnamefont
  {Ciocys}}, \bibinfo {author} {\bibfnamefont {Y.}~\bibnamefont {Wang}},
  \bibinfo {author} {\bibfnamefont {E.~S.}\ \bibnamefont {Choi}}, \bibinfo
  {author} {\bibfnamefont {R.}~\bibnamefont {Basnet}}, \bibinfo {author}
  {\bibfnamefont {A.}~\bibnamefont {Fereidouni}}, \emph {et~al.},\ }\bibfield
  {title} {\bibinfo {title} {Evidence for a magnetic-field-induced ideal
  type-ii weyl state in antiferromagnetic topological insulator mn (bi 1- x sb
  x) 2 te 4},\ }\href@noop {} {\bibfield  {journal} {\bibinfo  {journal}
  {Physical Review X}\ }\textbf {\bibinfo {volume} {11}},\ \bibinfo {pages}
  {031032} (\bibinfo {year} {2021})}\BibitemShut {NoStop}%
\bibitem [{\citenamefont {Yan}\ \emph {et~al.}(2019{\natexlab{a}})\citenamefont
  {Yan}, \citenamefont {Okamoto}, \citenamefont {McGuire}, \citenamefont {May},
  \citenamefont {McQueeney},\ and\ \citenamefont {Sales}}]{yan2019evolution}%
  \BibitemOpen
  \bibfield  {author} {\bibinfo {author} {\bibfnamefont {J.-Q.}\ \bibnamefont
  {Yan}}, \bibinfo {author} {\bibfnamefont {S.}~\bibnamefont {Okamoto}},
  \bibinfo {author} {\bibfnamefont {M.~A.}\ \bibnamefont {McGuire}}, \bibinfo
  {author} {\bibfnamefont {A.~F.}\ \bibnamefont {May}}, \bibinfo {author}
  {\bibfnamefont {R.~J.}\ \bibnamefont {McQueeney}},\ and\ \bibinfo {author}
  {\bibfnamefont {B.~C.}\ \bibnamefont {Sales}},\ }\bibfield  {title} {\bibinfo
  {title} {Evolution of structural, magnetic, and transport properties in mnbi
  2- x sb x te 4},\ }\href@noop {} {\bibfield  {journal} {\bibinfo  {journal}
  {Physical Review B}\ }\textbf {\bibinfo {volume} {100}},\ \bibinfo {pages}
  {104409} (\bibinfo {year} {2019}{\natexlab{a}})}\BibitemShut {NoStop}%
\bibitem [{\citenamefont {Li}\ \emph {et~al.}(2019{\natexlab{a}})\citenamefont
  {Li}, \citenamefont {Li}, \citenamefont {Du}, \citenamefont {Wang},
  \citenamefont {Gu}, \citenamefont {Zhang}, \citenamefont {He}, \citenamefont
  {Duan},\ and\ \citenamefont {Xu}}]{li2019intrinsic}%
  \BibitemOpen
  \bibfield  {author} {\bibinfo {author} {\bibfnamefont {J.}~\bibnamefont
  {Li}}, \bibinfo {author} {\bibfnamefont {Y.}~\bibnamefont {Li}}, \bibinfo
  {author} {\bibfnamefont {S.}~\bibnamefont {Du}}, \bibinfo {author}
  {\bibfnamefont {Z.}~\bibnamefont {Wang}}, \bibinfo {author} {\bibfnamefont
  {B.-L.}\ \bibnamefont {Gu}}, \bibinfo {author} {\bibfnamefont {S.-C.}\
  \bibnamefont {Zhang}}, \bibinfo {author} {\bibfnamefont {K.}~\bibnamefont
  {He}}, \bibinfo {author} {\bibfnamefont {W.}~\bibnamefont {Duan}},\ and\
  \bibinfo {author} {\bibfnamefont {Y.}~\bibnamefont {Xu}},\ }\bibfield
  {title} {\bibinfo {title} {Intrinsic magnetic topological insulators in van
  der waals layered mnbi2te4-family materials},\ }\href@noop {} {\bibfield
  {journal} {\bibinfo  {journal} {Science Advances}\ }\textbf {\bibinfo
  {volume} {5}},\ \bibinfo {pages} {eaaw5685} (\bibinfo {year}
  {2019}{\natexlab{a}})}\BibitemShut {NoStop}%
\bibitem [{\citenamefont {Petrov}\ \emph {et~al.}(2021)\citenamefont {Petrov},
  \citenamefont {Men'shov}, \citenamefont {Rusinov}, \citenamefont {Hoffmann},
  \citenamefont {Ernst}, \citenamefont {Otrokov}, \citenamefont {Dugaev},
  \citenamefont {Menshchikova~V},\ and\ \citenamefont
  {Chulkov~V}}]{petrov2021domain}%
  \BibitemOpen
  \bibfield  {author} {\bibinfo {author} {\bibfnamefont {E.}~\bibnamefont
  {Petrov}}, \bibinfo {author} {\bibfnamefont {V.}~\bibnamefont {Men'shov}},
  \bibinfo {author} {\bibfnamefont {I.}~\bibnamefont {Rusinov}}, \bibinfo
  {author} {\bibfnamefont {M.}~\bibnamefont {Hoffmann}}, \bibinfo {author}
  {\bibfnamefont {A.}~\bibnamefont {Ernst}}, \bibinfo {author} {\bibfnamefont
  {M.}~\bibnamefont {Otrokov}}, \bibinfo {author} {\bibfnamefont
  {V.}~\bibnamefont {Dugaev}}, \bibinfo {author} {\bibfnamefont
  {T.}~\bibnamefont {Menshchikova~V}},\ and\ \bibinfo {author} {\bibfnamefont
  {E.}~\bibnamefont {Chulkov~V}},\ }\bibfield  {title} {\bibinfo {title}
  {Domain wall induced spin-polarized flat bands in antiferromagnetic
  topological insulators},\ }\href@noop {} {\bibfield  {journal} {\bibinfo
  {journal} {Physical Review B}\ }\textbf {\bibinfo {volume} {103}} (\bibinfo
  {year} {2021})}\BibitemShut {NoStop}%
\bibitem [{\citenamefont {Zhang}\ \emph
  {et~al.}(2023{\natexlab{a}})\citenamefont {Zhang}, \citenamefont {Wang},
  \citenamefont {Yang}, \citenamefont {Zhang},\ and\ \citenamefont
  {Jia}}]{zhang2023strain}%
  \BibitemOpen
  \bibfield  {author} {\bibinfo {author} {\bibfnamefont {Y.}~\bibnamefont
  {Zhang}}, \bibinfo {author} {\bibfnamefont {Y.}~\bibnamefont {Wang}},
  \bibinfo {author} {\bibfnamefont {W.}~\bibnamefont {Yang}}, \bibinfo {author}
  {\bibfnamefont {H.}~\bibnamefont {Zhang}},\ and\ \bibinfo {author}
  {\bibfnamefont {J.}~\bibnamefont {Jia}},\ }\bibfield  {title} {\bibinfo
  {title} {Strain-tunable magnetism and topological states in layered vbi 2 te
  4},\ }\href@noop {} {\bibfield  {journal} {\bibinfo  {journal} {Physical
  Chemistry Chemical Physics}\ }\textbf {\bibinfo {volume} {25}},\ \bibinfo
  {pages} {28189} (\bibinfo {year} {2023}{\natexlab{a}})}\BibitemShut {NoStop}%
\bibitem [{\citenamefont {Altena}\ \emph {et~al.}(2023)\citenamefont {Altena},
  \citenamefont {Jansen}, \citenamefont {Tsvetanova},\ and\ \citenamefont
  {Brinkman}}]{altena2023phase}%
  \BibitemOpen
  \bibfield  {author} {\bibinfo {author} {\bibfnamefont {M.}~\bibnamefont
  {Altena}}, \bibinfo {author} {\bibfnamefont {T.}~\bibnamefont {Jansen}},
  \bibinfo {author} {\bibfnamefont {M.}~\bibnamefont {Tsvetanova}},\ and\
  \bibinfo {author} {\bibfnamefont {A.}~\bibnamefont {Brinkman}},\ }\bibfield
  {title} {\bibinfo {title} {Phase separation prevents the synthesis of vbi2te4
  by molecular beam epitaxy},\ }\href@noop {} {\bibfield  {journal} {\bibinfo
  {journal} {Nanomaterials}\ }\textbf {\bibinfo {volume} {14}},\ \bibinfo
  {pages} {87} (\bibinfo {year} {2023})}\BibitemShut {NoStop}%
\bibitem [{\citenamefont {Gonz\'alez-Hern\'andez}\ and\ \citenamefont
  {Uribe}(2024{\natexlab{a}})}]{PhysRevB.110.125129}%
  \BibitemOpen
  \bibfield  {author} {\bibinfo {author} {\bibfnamefont {R.}~\bibnamefont
  {Gonz\'alez-Hern\'andez}}\ and\ \bibinfo {author} {\bibfnamefont
  {B.}~\bibnamefont {Uribe}},\ }\bibfield  {title} {\bibinfo {title} {Average
  spin chern number},\ }\href {https://doi.org/10.1103/PhysRevB.110.125129}
  {\bibfield  {journal} {\bibinfo  {journal} {Phys. Rev. B}\ }\textbf {\bibinfo
  {volume} {110}},\ \bibinfo {pages} {125129} (\bibinfo {year}
  {2024}{\natexlab{a}})}\BibitemShut {NoStop}%
\bibitem [{\citenamefont {Kresse}\ and\ \citenamefont
  {Furthm{\"u}ller}(1993)}]{vasp}%
  \BibitemOpen
  \bibfield  {author} {\bibinfo {author} {\bibfnamefont {G.}~\bibnamefont
  {Kresse}}\ and\ \bibinfo {author} {\bibfnamefont {J.}~\bibnamefont
  {Furthm{\"u}ller}},\ }\bibfield  {title} {\bibinfo {title} {Vasp the guide
  (universit{\"a}t wien, wien, austria, 2007); g. kresse and j. hafner},\
  }\href@noop {} {\bibfield  {journal} {\bibinfo  {journal} {Phys. Rev. B}\
  }\textbf {\bibinfo {volume} {47}},\ \bibinfo {pages} {558} (\bibinfo {year}
  {1993})},\ \bibinfo {note} {en :
  \href{https://www.vasp.at/wiki}{www.vasp.at/wiki}}\BibitemShut {NoStop}%
\bibitem [{\citenamefont {Li}\ and\ \citenamefont {Li}(2022)}]{li2022}%
  \BibitemOpen
  \bibfield  {author} {\bibinfo {author} {\bibfnamefont {S.}~\bibnamefont
  {Li}}\ and\ \bibinfo {author} {\bibfnamefont {J.}~\bibnamefont {Li}},\
  }\href@noop {} {\emph {\bibinfo {title} {Introduction to Computational
  Nanomechanics}}}\ (\bibinfo  {publisher} {Cambridge University Press},\
  \bibinfo {year} {2022})\BibitemShut {NoStop}%
\bibitem [{\citenamefont {Dudarev}\ \emph {et~al.}(1998)\citenamefont
  {Dudarev}, \citenamefont {Botton}, \citenamefont {Savrasov}, \citenamefont
  {Humphreys},\ and\ \citenamefont {Sutton}}]{dudarev1998}%
  \BibitemOpen
  \bibfield  {author} {\bibinfo {author} {\bibfnamefont {S.~L.}\ \bibnamefont
  {Dudarev}}, \bibinfo {author} {\bibfnamefont {G.~A.}\ \bibnamefont {Botton}},
  \bibinfo {author} {\bibfnamefont {S.~Y.}\ \bibnamefont {Savrasov}}, \bibinfo
  {author} {\bibfnamefont {C.}~\bibnamefont {Humphreys}},\ and\ \bibinfo
  {author} {\bibfnamefont {A.~P.}\ \bibnamefont {Sutton}},\ }\bibfield  {title}
  {\bibinfo {title} {Electron-energy-loss spectra and the structural stability
  of nickel oxide: An lsda+ u study},\ }\href@noop {} {\bibfield  {journal}
  {\bibinfo  {journal} {Physical Review B}\ }\textbf {\bibinfo {volume} {57}},\
  \bibinfo {pages} {1505} (\bibinfo {year} {1998})}\BibitemShut {NoStop}%
\bibitem [{\citenamefont {Cococcioni}\ and\ \citenamefont
  {De~Gironcoli}(2005)}]{cococcioni2005linear}%
  \BibitemOpen
  \bibfield  {author} {\bibinfo {author} {\bibfnamefont {M.}~\bibnamefont
  {Cococcioni}}\ and\ \bibinfo {author} {\bibfnamefont {S.}~\bibnamefont
  {De~Gironcoli}},\ }\bibfield  {title} {\bibinfo {title} {Linear response
  approach to the calculation of the effective interaction parameters in the
  lda+ u method},\ }\href@noop {} {\bibfield  {journal} {\bibinfo  {journal}
  {Physical Review B—Condensed Matter and Materials Physics}\ }\textbf
  {\bibinfo {volume} {71}},\ \bibinfo {pages} {035105} (\bibinfo {year}
  {2005})}\BibitemShut {NoStop}%
\bibitem [{\citenamefont {Zhou}\ \emph {et~al.}(2004)\citenamefont {Zhou},
  \citenamefont {Cococcioni}, \citenamefont {Marianetti}, \citenamefont
  {Morgan},\ and\ \citenamefont {Ceder}}]{zhou2004first}%
  \BibitemOpen
  \bibfield  {author} {\bibinfo {author} {\bibfnamefont {F.}~\bibnamefont
  {Zhou}}, \bibinfo {author} {\bibfnamefont {M.}~\bibnamefont {Cococcioni}},
  \bibinfo {author} {\bibfnamefont {C.~A.}\ \bibnamefont {Marianetti}},
  \bibinfo {author} {\bibfnamefont {D.}~\bibnamefont {Morgan}},\ and\ \bibinfo
  {author} {\bibfnamefont {G.}~\bibnamefont {Ceder}},\ }\bibfield  {title}
  {\bibinfo {title} {First-principles prediction of redox potentials in
  transition-metal compounds with lda+ u},\ }\href@noop {} {\bibfield
  {journal} {\bibinfo  {journal} {Physical Review B—Condensed Matter and
  Materials Physics}\ }\textbf {\bibinfo {volume} {70}},\ \bibinfo {pages}
  {235121} (\bibinfo {year} {2004})}\BibitemShut {NoStop}%
\bibitem [{\citenamefont {Wang}\ and\ \citenamefont
  {Navrotsky}(2004)}]{wang2004enthalpy}%
  \BibitemOpen
  \bibfield  {author} {\bibinfo {author} {\bibfnamefont {M.}~\bibnamefont
  {Wang}}\ and\ \bibinfo {author} {\bibfnamefont {A.}~\bibnamefont
  {Navrotsky}},\ }\bibfield  {title} {\bibinfo {title} {Enthalpy of formation
  of linio2, licoo2 and their solid solution, lini1- xcoxo2},\ }\href@noop {}
  {\bibfield  {journal} {\bibinfo  {journal} {Solid State Ionics}\ }\textbf
  {\bibinfo {volume} {166}},\ \bibinfo {pages} {167} (\bibinfo {year}
  {2004})}\BibitemShut {NoStop}%
\bibitem [{\citenamefont {Grimme}\ \emph {et~al.}(2010)\citenamefont {Grimme},
  \citenamefont {Antony}, \citenamefont {Ehrlich},\ and\ \citenamefont
  {Krieg}}]{grimme2010consistent}%
  \BibitemOpen
  \bibfield  {author} {\bibinfo {author} {\bibfnamefont {S.}~\bibnamefont
  {Grimme}}, \bibinfo {author} {\bibfnamefont {J.}~\bibnamefont {Antony}},
  \bibinfo {author} {\bibfnamefont {S.}~\bibnamefont {Ehrlich}},\ and\ \bibinfo
  {author} {\bibfnamefont {H.}~\bibnamefont {Krieg}},\ }\bibfield  {title}
  {\bibinfo {title} {A consistent and accurate ab initio parametrization of
  density functional dispersion correction (dft-d) for the 94 elements h-pu},\
  }\href@noop {} {\bibfield  {journal} {\bibinfo  {journal} {The Journal of
  chemical physics}\ }\textbf {\bibinfo {volume} {132}} (\bibinfo {year}
  {2010})}\BibitemShut {NoStop}%
\bibitem [{\citenamefont {Khalaf}\ \emph {et~al.}(2018)\citenamefont {Khalaf},
  \citenamefont {Po}, \citenamefont {Vishwanath},\ and\ \citenamefont
  {Watanabe}}]{khalaf2018symmetry}%
  \BibitemOpen
  \bibfield  {author} {\bibinfo {author} {\bibfnamefont {E.}~\bibnamefont
  {Khalaf}}, \bibinfo {author} {\bibfnamefont {H.~C.}\ \bibnamefont {Po}},
  \bibinfo {author} {\bibfnamefont {A.}~\bibnamefont {Vishwanath}},\ and\
  \bibinfo {author} {\bibfnamefont {H.}~\bibnamefont {Watanabe}},\ }\bibfield
  {title} {\bibinfo {title} {Symmetry indicators and anomalous surface states
  of topological crystalline insulators},\ }\href@noop {} {\bibfield  {journal}
  {\bibinfo  {journal} {Physical Review X}\ }\textbf {\bibinfo {volume} {8}},\
  \bibinfo {pages} {031070} (\bibinfo {year} {2018})}\BibitemShut {NoStop}%
\bibitem [{\citenamefont {Song}\ \emph {et~al.}(2018)\citenamefont {Song},
  \citenamefont {Zhang}, \citenamefont {Fang},\ and\ \citenamefont
  {Fang}}]{song2018quantitative}%
  \BibitemOpen
  \bibfield  {author} {\bibinfo {author} {\bibfnamefont {Z.}~\bibnamefont
  {Song}}, \bibinfo {author} {\bibfnamefont {T.}~\bibnamefont {Zhang}},
  \bibinfo {author} {\bibfnamefont {Z.}~\bibnamefont {Fang}},\ and\ \bibinfo
  {author} {\bibfnamefont {C.}~\bibnamefont {Fang}},\ }\bibfield  {title}
  {\bibinfo {title} {Quantitative mappings between symmetry and topology in
  solids},\ }\href@noop {} {\bibfield  {journal} {\bibinfo  {journal} {Nature
  communications}\ }\textbf {\bibinfo {volume} {9}},\ \bibinfo {pages} {3530}
  (\bibinfo {year} {2018})}\BibitemShut {NoStop}%
\bibitem [{\citenamefont {Tanaka}\ \emph {et~al.}(2020)\citenamefont {Tanaka},
  \citenamefont {Takahashi}, \citenamefont {Zhang},\ and\ \citenamefont
  {Murakami}}]{tanaka2020theory}%
  \BibitemOpen
  \bibfield  {author} {\bibinfo {author} {\bibfnamefont {Y.}~\bibnamefont
  {Tanaka}}, \bibinfo {author} {\bibfnamefont {R.}~\bibnamefont {Takahashi}},
  \bibinfo {author} {\bibfnamefont {T.}~\bibnamefont {Zhang}},\ and\ \bibinfo
  {author} {\bibfnamefont {S.}~\bibnamefont {Murakami}},\ }\bibfield  {title}
  {\bibinfo {title} {Theory of inversion-z 4 protected topological chiral hinge
  states and its applications to layered antiferromagnets},\ }\href@noop {}
  {\bibfield  {journal} {\bibinfo  {journal} {Physical Review Research}\
  }\textbf {\bibinfo {volume} {2}},\ \bibinfo {pages} {043274} (\bibinfo {year}
  {2020})}\BibitemShut {NoStop}%
\bibitem [{\citenamefont {Wang}\ \emph {et~al.}(2019)\citenamefont {Wang},
  \citenamefont {Wieder}, \citenamefont {Li}, \citenamefont {Yan},\ and\
  \citenamefont {Bernevig}}]{wang2019higher}%
  \BibitemOpen
  \bibfield  {author} {\bibinfo {author} {\bibfnamefont {Z.}~\bibnamefont
  {Wang}}, \bibinfo {author} {\bibfnamefont {B.~J.}\ \bibnamefont {Wieder}},
  \bibinfo {author} {\bibfnamefont {J.}~\bibnamefont {Li}}, \bibinfo {author}
  {\bibfnamefont {B.}~\bibnamefont {Yan}},\ and\ \bibinfo {author}
  {\bibfnamefont {B.~A.}\ \bibnamefont {Bernevig}},\ }\bibfield  {title}
  {\bibinfo {title} {Higher-order topology, monopole nodal lines, and the
  origin of large fermi arcs in transition metal dichalcogenides x te 2 (x= mo,
  w)},\ }\href@noop {} {\bibfield  {journal} {\bibinfo  {journal} {Physical
  review letters}\ }\textbf {\bibinfo {volume} {123}},\ \bibinfo {pages}
  {186401} (\bibinfo {year} {2019})}\BibitemShut {NoStop}%
\bibitem [{\citenamefont {Mostofi}\ \emph {et~al.}(2008)\citenamefont
  {Mostofi}, \citenamefont {Yates}, \citenamefont {Lee}, \citenamefont {Souza},
  \citenamefont {Vanderbilt},\ and\ \citenamefont
  {Marzari}}]{mostofi2008wannier90}%
  \BibitemOpen
  \bibfield  {author} {\bibinfo {author} {\bibfnamefont {A.~A.}\ \bibnamefont
  {Mostofi}}, \bibinfo {author} {\bibfnamefont {J.~R.}\ \bibnamefont {Yates}},
  \bibinfo {author} {\bibfnamefont {Y.-S.}\ \bibnamefont {Lee}}, \bibinfo
  {author} {\bibfnamefont {I.}~\bibnamefont {Souza}}, \bibinfo {author}
  {\bibfnamefont {D.}~\bibnamefont {Vanderbilt}},\ and\ \bibinfo {author}
  {\bibfnamefont {N.}~\bibnamefont {Marzari}},\ }\bibfield  {title} {\bibinfo
  {title} {wannier90: A tool for obtaining maximally-localised wannier
  functions},\ }\href@noop {} {\bibfield  {journal} {\bibinfo  {journal}
  {Computer physics communications}\ }\textbf {\bibinfo {volume} {178}},\
  \bibinfo {pages} {685} (\bibinfo {year} {2008})}\BibitemShut {NoStop}%
\bibitem [{\citenamefont {Wu}\ \emph {et~al.}(2018)\citenamefont {Wu},
  \citenamefont {Zhang}, \citenamefont {Song}, \citenamefont {Troyer},\ and\
  \citenamefont {Soluyanov}}]{WU2017}%
  \BibitemOpen
  \bibfield  {author} {\bibinfo {author} {\bibfnamefont {Q.}~\bibnamefont
  {Wu}}, \bibinfo {author} {\bibfnamefont {S.}~\bibnamefont {Zhang}}, \bibinfo
  {author} {\bibfnamefont {H.-F.}\ \bibnamefont {Song}}, \bibinfo {author}
  {\bibfnamefont {M.}~\bibnamefont {Troyer}},\ and\ \bibinfo {author}
  {\bibfnamefont {A.~A.}\ \bibnamefont {Soluyanov}},\ }\bibfield  {title}
  {\bibinfo {title} {Wanniertools : An open-source software package for novel
  topological materials},\ }\href
  {https://doi.org/https://doi.org/10.1016/j.cpc.2017.09.033} {\bibfield
  {journal} {\bibinfo  {journal} {Computer Physics Communications}\ }\textbf
  {\bibinfo {volume} {224}},\ \bibinfo {pages} {405 } (\bibinfo {year}
  {2018})}\BibitemShut {NoStop}%
\bibitem [{\citenamefont {Yan}\ \emph {et~al.}(2019{\natexlab{b}})\citenamefont
  {Yan}, \citenamefont {Zhang}, \citenamefont {Heitmann}, \citenamefont
  {Huang}, \citenamefont {Chen}, \citenamefont {Cheng}, \citenamefont {Wu},
  \citenamefont {Vaknin}, \citenamefont {Sales},\ and\ \citenamefont
  {McQueeney}}]{yan2019crystal}%
  \BibitemOpen
  \bibfield  {author} {\bibinfo {author} {\bibfnamefont {J.-Q.}\ \bibnamefont
  {Yan}}, \bibinfo {author} {\bibfnamefont {Q.}~\bibnamefont {Zhang}}, \bibinfo
  {author} {\bibfnamefont {T.}~\bibnamefont {Heitmann}}, \bibinfo {author}
  {\bibfnamefont {Z.}~\bibnamefont {Huang}}, \bibinfo {author} {\bibfnamefont
  {K.}~\bibnamefont {Chen}}, \bibinfo {author} {\bibfnamefont {J.-G.}\
  \bibnamefont {Cheng}}, \bibinfo {author} {\bibfnamefont {W.}~\bibnamefont
  {Wu}}, \bibinfo {author} {\bibfnamefont {D.}~\bibnamefont {Vaknin}}, \bibinfo
  {author} {\bibfnamefont {B.~C.}\ \bibnamefont {Sales}},\ and\ \bibinfo
  {author} {\bibfnamefont {R.~J.}\ \bibnamefont {McQueeney}},\ }\bibfield
  {title} {\bibinfo {title} {Crystal growth and magnetic structure of mnbi 2 te
  4},\ }\href@noop {} {\bibfield  {journal} {\bibinfo  {journal} {Physical
  Review Materials}\ }\textbf {\bibinfo {volume} {3}},\ \bibinfo {pages}
  {064202} (\bibinfo {year} {2019}{\natexlab{b}})}\BibitemShut {NoStop}%
\bibitem [{\citenamefont {Ding}\ \emph {et~al.}(2020)\citenamefont {Ding},
  \citenamefont {Hu}, \citenamefont {Ye}, \citenamefont {Feng}, \citenamefont
  {Ni},\ and\ \citenamefont {Cao}}]{ding2020crystal}%
  \BibitemOpen
  \bibfield  {author} {\bibinfo {author} {\bibfnamefont {L.}~\bibnamefont
  {Ding}}, \bibinfo {author} {\bibfnamefont {C.}~\bibnamefont {Hu}}, \bibinfo
  {author} {\bibfnamefont {F.}~\bibnamefont {Ye}}, \bibinfo {author}
  {\bibfnamefont {E.}~\bibnamefont {Feng}}, \bibinfo {author} {\bibfnamefont
  {N.}~\bibnamefont {Ni}},\ and\ \bibinfo {author} {\bibfnamefont
  {H.}~\bibnamefont {Cao}},\ }\bibfield  {title} {\bibinfo {title} {Crystal and
  magnetic structures of magnetic topological insulators mnbi 2 te 4 and mnbi 4
  te 7},\ }\href@noop {} {\bibfield  {journal} {\bibinfo  {journal} {Physical
  Review B}\ }\textbf {\bibinfo {volume} {101}},\ \bibinfo {pages} {020412}
  (\bibinfo {year} {2020})}\BibitemShut {NoStop}%
\bibitem [{\citenamefont {Stokes}\ and\ \citenamefont {Hatch}(2005)}]{FIND}%
  \BibitemOpen
  \bibfield  {author} {\bibinfo {author} {\bibfnamefont {H.~T.}\ \bibnamefont
  {Stokes}}\ and\ \bibinfo {author} {\bibfnamefont {D.~M.}\ \bibnamefont
  {Hatch}},\ }\bibfield  {title} {\bibinfo {title} {Findsym: program for
  identifying the space-group symmetry of a crystal},\ }\href@noop {}
  {\bibfield  {journal} {\bibinfo  {journal} {Journal of Applied
  Crystallography}\ }\textbf {\bibinfo {volume} {38}},\ \bibinfo {pages} {237}
  (\bibinfo {year} {2005})}\BibitemShut {NoStop}%
\bibitem [{\citenamefont {Gonz\'alez-Hern\'andez}\ and\ \citenamefont
  {Uribe}(2024{\natexlab{b}})}]{spinweyl}%
  \BibitemOpen
  \bibfield  {author} {\bibinfo {author} {\bibfnamefont {R.}~\bibnamefont
  {Gonz\'alez-Hern\'andez}}\ and\ \bibinfo {author} {\bibfnamefont
  {B.}~\bibnamefont {Uribe}},\ }\bibfield  {title} {\bibinfo {title} {Spin weyl
  topological insulators},\ }\href
  {https://doi.org/10.1103/PhysRevB.109.045126} {\bibfield  {journal} {\bibinfo
   {journal} {Phys. Rev. B}\ }\textbf {\bibinfo {volume} {109}},\ \bibinfo
  {pages} {045126} (\bibinfo {year} {2024}{\natexlab{b}})}\BibitemShut
  {NoStop}%
\bibitem [{\citenamefont {Li}\ \emph {et~al.}(2019{\natexlab{b}})\citenamefont
  {Li}, \citenamefont {Wang}, \citenamefont {Zhang}, \citenamefont {Gu},
  \citenamefont {Duan},\ and\ \citenamefont {Xu}}]{li2019magnetically}%
  \BibitemOpen
  \bibfield  {author} {\bibinfo {author} {\bibfnamefont {J.}~\bibnamefont
  {Li}}, \bibinfo {author} {\bibfnamefont {C.}~\bibnamefont {Wang}}, \bibinfo
  {author} {\bibfnamefont {Z.}~\bibnamefont {Zhang}}, \bibinfo {author}
  {\bibfnamefont {B.-L.}\ \bibnamefont {Gu}}, \bibinfo {author} {\bibfnamefont
  {W.}~\bibnamefont {Duan}},\ and\ \bibinfo {author} {\bibfnamefont
  {Y.}~\bibnamefont {Xu}},\ }\bibfield  {title} {\bibinfo {title} {Magnetically
  controllable topological quantum phase transitions in the antiferromagnetic
  topological insulator mnbi 2 te 4},\ }\href@noop {} {\bibfield  {journal}
  {\bibinfo  {journal} {Physical Review B}\ }\textbf {\bibinfo {volume}
  {100}},\ \bibinfo {pages} {121103} (\bibinfo {year}
  {2019}{\natexlab{b}})}\BibitemShut {NoStop}%
\bibitem [{\citenamefont {Hao}\ \emph {et~al.}(2019)\citenamefont {Hao},
  \citenamefont {Liu}, \citenamefont {Feng}, \citenamefont {Ma}, \citenamefont
  {Schwier}, \citenamefont {Arita}, \citenamefont {Kumar}, \citenamefont {Hu},
  \citenamefont {Lu}, \citenamefont {Zeng} \emph {et~al.}}]{hao2019gapless}%
  \BibitemOpen
  \bibfield  {author} {\bibinfo {author} {\bibfnamefont {Y.-J.}\ \bibnamefont
  {Hao}}, \bibinfo {author} {\bibfnamefont {P.}~\bibnamefont {Liu}}, \bibinfo
  {author} {\bibfnamefont {Y.}~\bibnamefont {Feng}}, \bibinfo {author}
  {\bibfnamefont {X.-M.}\ \bibnamefont {Ma}}, \bibinfo {author} {\bibfnamefont
  {E.~F.}\ \bibnamefont {Schwier}}, \bibinfo {author} {\bibfnamefont
  {M.}~\bibnamefont {Arita}}, \bibinfo {author} {\bibfnamefont
  {S.}~\bibnamefont {Kumar}}, \bibinfo {author} {\bibfnamefont
  {C.}~\bibnamefont {Hu}}, \bibinfo {author} {\bibfnamefont {R.}~\bibnamefont
  {Lu}}, \bibinfo {author} {\bibfnamefont {M.}~\bibnamefont {Zeng}}, \emph
  {et~al.},\ }\bibfield  {title} {\bibinfo {title} {Gapless surface dirac cone
  in antiferromagnetic topological insulator mnbi 2 te 4},\ }\href@noop {}
  {\bibfield  {journal} {\bibinfo  {journal} {Physical Review X}\ }\textbf
  {\bibinfo {volume} {9}},\ \bibinfo {pages} {041038} (\bibinfo {year}
  {2019})}\BibitemShut {NoStop}%
\bibitem [{\citenamefont {Chen}\ \emph {et~al.}(2019)\citenamefont {Chen},
  \citenamefont {Xu}, \citenamefont {Li}, \citenamefont {Li}, \citenamefont
  {Wang}, \citenamefont {Zhang}, \citenamefont {Li}, \citenamefont {Wu},
  \citenamefont {Liang}, \citenamefont {Chen} \emph
  {et~al.}}]{chen2019surface}%
  \BibitemOpen
  \bibfield  {author} {\bibinfo {author} {\bibfnamefont {Y.}~\bibnamefont
  {Chen}}, \bibinfo {author} {\bibfnamefont {L.}~\bibnamefont {Xu}}, \bibinfo
  {author} {\bibfnamefont {J.}~\bibnamefont {Li}}, \bibinfo {author}
  {\bibfnamefont {Y.}~\bibnamefont {Li}}, \bibinfo {author} {\bibfnamefont
  {H.}~\bibnamefont {Wang}}, \bibinfo {author} {\bibfnamefont {C.}~\bibnamefont
  {Zhang}}, \bibinfo {author} {\bibfnamefont {H.}~\bibnamefont {Li}}, \bibinfo
  {author} {\bibfnamefont {Y.}~\bibnamefont {Wu}}, \bibinfo {author}
  {\bibfnamefont {A.}~\bibnamefont {Liang}}, \bibinfo {author} {\bibfnamefont
  {C.}~\bibnamefont {Chen}}, \emph {et~al.},\ }\bibfield  {title} {\bibinfo
  {title} {Topological electronic structure and its temperature evolution in
  antiferromagnetic topological insulator mnbi 2 te 4 and supplementary
  material si ii. ab-initio calculation of topological surface states},\
  }\href@noop {} {\bibfield  {journal} {\bibinfo  {journal} {Physical Review
  X}\ }\textbf {\bibinfo {volume} {9}},\ \bibinfo {pages} {041040} (\bibinfo
  {year} {2019})}\BibitemShut {NoStop}%
\bibitem [{\citenamefont {Sun}\ \emph {et~al.}(2020)\citenamefont {Sun},
  \citenamefont {Wang}, \citenamefont {Zhang}, \citenamefont {Chen},
  \citenamefont {Zhao}, \citenamefont {Liu}, \citenamefont {Liu}, \citenamefont
  {Chen}, \citenamefont {Lu},\ and\ \citenamefont {Xie}}]{PhysRevB.102.241406}%
  \BibitemOpen
  \bibfield  {author} {\bibinfo {author} {\bibfnamefont {H.-P.}\ \bibnamefont
  {Sun}}, \bibinfo {author} {\bibfnamefont {C.~M.}\ \bibnamefont {Wang}},
  \bibinfo {author} {\bibfnamefont {S.-B.}\ \bibnamefont {Zhang}}, \bibinfo
  {author} {\bibfnamefont {R.}~\bibnamefont {Chen}}, \bibinfo {author}
  {\bibfnamefont {Y.}~\bibnamefont {Zhao}}, \bibinfo {author} {\bibfnamefont
  {C.}~\bibnamefont {Liu}}, \bibinfo {author} {\bibfnamefont {Q.}~\bibnamefont
  {Liu}}, \bibinfo {author} {\bibfnamefont {C.}~\bibnamefont {Chen}}, \bibinfo
  {author} {\bibfnamefont {H.-Z.}\ \bibnamefont {Lu}},\ and\ \bibinfo {author}
  {\bibfnamefont {X.~C.}\ \bibnamefont {Xie}},\ }\bibfield  {title} {\bibinfo
  {title} {Analytical solution for the surface states of the antiferromagnetic
  topological insulator ${\mathrm{mnbi}}_{2}{\mathrm{te}}_{4}$},\ }\href
  {https://doi.org/10.1103/PhysRevB.102.241406} {\bibfield  {journal} {\bibinfo
   {journal} {Phys. Rev. B}\ }\textbf {\bibinfo {volume} {102}},\ \bibinfo
  {pages} {241406} (\bibinfo {year} {2020})}\BibitemShut {NoStop}%
\bibitem [{\citenamefont {Bernevig}\ \emph {et~al.}(2022)\citenamefont
  {Bernevig}, \citenamefont {Felser},\ and\ \citenamefont
  {Beidenkopf}}]{bernevig2022progress}%
  \BibitemOpen
  \bibfield  {author} {\bibinfo {author} {\bibfnamefont {B.~A.}\ \bibnamefont
  {Bernevig}}, \bibinfo {author} {\bibfnamefont {C.}~\bibnamefont {Felser}},\
  and\ \bibinfo {author} {\bibfnamefont {H.}~\bibnamefont {Beidenkopf}},\
  }\bibfield  {title} {\bibinfo {title} {Progress and prospects in magnetic
  topological materials},\ }\href@noop {} {\bibfield  {journal} {\bibinfo
  {journal} {Nature}\ }\textbf {\bibinfo {volume} {603}},\ \bibinfo {pages}
  {41} (\bibinfo {year} {2022})}\BibitemShut {NoStop}%
\bibitem [{\citenamefont {Ernzerhof}\ and\ \citenamefont
  {Scuseria}(1999)}]{ernzerhof1999}%
  \BibitemOpen
  \bibfield  {author} {\bibinfo {author} {\bibfnamefont {M.}~\bibnamefont
  {Ernzerhof}}\ and\ \bibinfo {author} {\bibfnamefont {G.~E.}\ \bibnamefont
  {Scuseria}},\ }\bibfield  {title} {\bibinfo {title} {Assessment of the
  perdew--burke--ernzerhof exchange-correlation functional},\ }\href@noop {}
  {\bibfield  {journal} {\bibinfo  {journal} {The Journal of chemical physics}\
  }\textbf {\bibinfo {volume} {110}},\ \bibinfo {pages} {5029} (\bibinfo {year}
  {1999})}\BibitemShut {NoStop}%
\bibitem [{\citenamefont {Zhang}\ \emph
  {et~al.}(2023{\natexlab{b}})\citenamefont {Zhang}, \citenamefont {Wang},
  \citenamefont {Yang}, \citenamefont {Zhang},\ and\ \citenamefont
  {Jia}}]{Zhang2023}%
  \BibitemOpen
  \bibfield  {author} {\bibinfo {author} {\bibfnamefont {Y.}~\bibnamefont
  {Zhang}}, \bibinfo {author} {\bibfnamefont {Y.}~\bibnamefont {Wang}},
  \bibinfo {author} {\bibfnamefont {W.}~\bibnamefont {Yang}}, \bibinfo {author}
  {\bibfnamefont {H.}~\bibnamefont {Zhang}},\ and\ \bibinfo {author}
  {\bibfnamefont {J.}~\bibnamefont {Jia}},\ }\bibfield  {title} {\bibinfo
  {title} {Strain-tunable magnetism and topological states in layered
  vbi2te4},\ }\href {https://doi.org/10.1039/D3CP03866A} {\bibfield  {journal}
  {\bibinfo  {journal} {Phys. Chem. Chem. Phys.}\ ,\ } (\bibinfo {year}
  {2023}{\natexlab{b}})}\BibitemShut {NoStop}%
\bibitem [{\citenamefont {Iraola}\ \emph {et~al.}(2022)\citenamefont {Iraola},
  \citenamefont {Ma{\~n}es}, \citenamefont {Bradlyn}, \citenamefont {Horton},
  \citenamefont {Neupert}, \citenamefont {Vergniory},\ and\ \citenamefont
  {Tsirkin}}]{iraola2022irrep}%
  \BibitemOpen
  \bibfield  {author} {\bibinfo {author} {\bibfnamefont {M.}~\bibnamefont
  {Iraola}}, \bibinfo {author} {\bibfnamefont {J.~L.}\ \bibnamefont
  {Ma{\~n}es}}, \bibinfo {author} {\bibfnamefont {B.}~\bibnamefont {Bradlyn}},
  \bibinfo {author} {\bibfnamefont {M.~K.}\ \bibnamefont {Horton}}, \bibinfo
  {author} {\bibfnamefont {T.}~\bibnamefont {Neupert}}, \bibinfo {author}
  {\bibfnamefont {M.~G.}\ \bibnamefont {Vergniory}},\ and\ \bibinfo {author}
  {\bibfnamefont {S.~S.}\ \bibnamefont {Tsirkin}},\ }\bibfield  {title}
  {\bibinfo {title} {Irrep: Symmetry eigenvalues and irreducible
  representations of ab initio band structures},\ }\href@noop {} {\bibfield
  {journal} {\bibinfo  {journal} {Computer Physics Communications}\ }\textbf
  {\bibinfo {volume} {272}},\ \bibinfo {pages} {108226} (\bibinfo {year}
  {2022})}\BibitemShut {NoStop}%
\bibitem [{\citenamefont {Zhang}\ \emph {et~al.}(2019)\citenamefont {Zhang},
  \citenamefont {Shi}, \citenamefont {Zhu}, \citenamefont {Xing}, \citenamefont
  {Zhang},\ and\ \citenamefont {Wang}}]{zhang2019topological}%
  \BibitemOpen
  \bibfield  {author} {\bibinfo {author} {\bibfnamefont {D.}~\bibnamefont
  {Zhang}}, \bibinfo {author} {\bibfnamefont {M.}~\bibnamefont {Shi}}, \bibinfo
  {author} {\bibfnamefont {T.}~\bibnamefont {Zhu}}, \bibinfo {author}
  {\bibfnamefont {D.}~\bibnamefont {Xing}}, \bibinfo {author} {\bibfnamefont
  {H.}~\bibnamefont {Zhang}},\ and\ \bibinfo {author} {\bibfnamefont
  {J.}~\bibnamefont {Wang}},\ }\bibfield  {title} {\bibinfo {title}
  {Topological axion states in the magnetic insulator mnbi 2 te 4 with the
  quantized magnetoelectric effect},\ }\href@noop {} {\bibfield  {journal}
  {\bibinfo  {journal} {Physical review letters}\ }\textbf {\bibinfo {volume}
  {122}},\ \bibinfo {pages} {206401} (\bibinfo {year} {2019})}\BibitemShut
  {NoStop}%
\bibitem [{\citenamefont {Li}\ \emph {et~al.}(2020)\citenamefont {Li},
  \citenamefont {Li}, \citenamefont {He}, \citenamefont {Wan}, \citenamefont
  {Duan},\ and\ \citenamefont {Xu}}]{li2020tunable}%
  \BibitemOpen
  \bibfield  {author} {\bibinfo {author} {\bibfnamefont {Z.}~\bibnamefont
  {Li}}, \bibinfo {author} {\bibfnamefont {J.}~\bibnamefont {Li}}, \bibinfo
  {author} {\bibfnamefont {K.}~\bibnamefont {He}}, \bibinfo {author}
  {\bibfnamefont {X.}~\bibnamefont {Wan}}, \bibinfo {author} {\bibfnamefont
  {W.}~\bibnamefont {Duan}},\ and\ \bibinfo {author} {\bibfnamefont
  {Y.}~\bibnamefont {Xu}},\ }\bibfield  {title} {\bibinfo {title} {Tunable
  interlayer magnetism and band topology in van der waals heterostructures of
  mn bi 2 te 4-family materials},\ }\href@noop {} {\bibfield  {journal}
  {\bibinfo  {journal} {Physical Review B}\ }\textbf {\bibinfo {volume}
  {102}},\ \bibinfo {pages} {081107} (\bibinfo {year} {2020})}\BibitemShut
  {NoStop}%
\bibitem [{\citenamefont {Liu}\ \emph {et~al.}(2020)\citenamefont {Liu},
  \citenamefont {Wang}, \citenamefont {Li}, \citenamefont {Wu}, \citenamefont
  {Li}, \citenamefont {Li}, \citenamefont {He}, \citenamefont {Xu},
  \citenamefont {Zhang},\ and\ \citenamefont {Wang}}]{liu2020robust}%
  \BibitemOpen
  \bibfield  {author} {\bibinfo {author} {\bibfnamefont {C.}~\bibnamefont
  {Liu}}, \bibinfo {author} {\bibfnamefont {Y.}~\bibnamefont {Wang}}, \bibinfo
  {author} {\bibfnamefont {H.}~\bibnamefont {Li}}, \bibinfo {author}
  {\bibfnamefont {Y.}~\bibnamefont {Wu}}, \bibinfo {author} {\bibfnamefont
  {Y.}~\bibnamefont {Li}}, \bibinfo {author} {\bibfnamefont {J.}~\bibnamefont
  {Li}}, \bibinfo {author} {\bibfnamefont {K.}~\bibnamefont {He}}, \bibinfo
  {author} {\bibfnamefont {Y.}~\bibnamefont {Xu}}, \bibinfo {author}
  {\bibfnamefont {J.}~\bibnamefont {Zhang}},\ and\ \bibinfo {author}
  {\bibfnamefont {Y.}~\bibnamefont {Wang}},\ }\bibfield  {title} {\bibinfo
  {title} {Robust axion insulator and chern insulator phases in a
  two-dimensional antiferromagnetic topological insulator},\ }\href@noop {}
  {\bibfield  {journal} {\bibinfo  {journal} {Nature materials}\ }\textbf
  {\bibinfo {volume} {19}},\ \bibinfo {pages} {522} (\bibinfo {year}
  {2020})}\BibitemShut {NoStop}%
\bibitem [{\citenamefont {Deng}\ \emph {et~al.}(2020)\citenamefont {Deng},
  \citenamefont {Yu}, \citenamefont {Shi}, \citenamefont {Guo}, \citenamefont
  {Xu}, \citenamefont {Wang}, \citenamefont {Chen},\ and\ \citenamefont
  {Zhang}}]{deng2020quantum}%
  \BibitemOpen
  \bibfield  {author} {\bibinfo {author} {\bibfnamefont {Y.}~\bibnamefont
  {Deng}}, \bibinfo {author} {\bibfnamefont {Y.}~\bibnamefont {Yu}}, \bibinfo
  {author} {\bibfnamefont {M.~Z.}\ \bibnamefont {Shi}}, \bibinfo {author}
  {\bibfnamefont {Z.}~\bibnamefont {Guo}}, \bibinfo {author} {\bibfnamefont
  {Z.}~\bibnamefont {Xu}}, \bibinfo {author} {\bibfnamefont {J.}~\bibnamefont
  {Wang}}, \bibinfo {author} {\bibfnamefont {X.~H.}\ \bibnamefont {Chen}},\
  and\ \bibinfo {author} {\bibfnamefont {Y.}~\bibnamefont {Zhang}},\ }\bibfield
   {title} {\bibinfo {title} {Quantum anomalous hall effect in intrinsic
  magnetic topological insulator mnbi2te4},\ }\href@noop {} {\bibfield
  {journal} {\bibinfo  {journal} {Science}\ }\textbf {\bibinfo {volume}
  {367}},\ \bibinfo {pages} {895} (\bibinfo {year} {2020})}\BibitemShut
  {NoStop}%
\bibitem [{\citenamefont {Rani}\ \emph {et~al.}(2019)\citenamefont {Rani},
  \citenamefont {Saxena}, \citenamefont {Sultana}, \citenamefont {Nagpal},
  \citenamefont {Islam}, \citenamefont {Patnaik},\ and\ \citenamefont
  {Awana}}]{rani2019crystal}%
  \BibitemOpen
  \bibfield  {author} {\bibinfo {author} {\bibfnamefont {P.}~\bibnamefont
  {Rani}}, \bibinfo {author} {\bibfnamefont {A.}~\bibnamefont {Saxena}},
  \bibinfo {author} {\bibfnamefont {R.}~\bibnamefont {Sultana}}, \bibinfo
  {author} {\bibfnamefont {V.}~\bibnamefont {Nagpal}}, \bibinfo {author}
  {\bibfnamefont {S.}~\bibnamefont {Islam}}, \bibinfo {author} {\bibfnamefont
  {S.}~\bibnamefont {Patnaik}},\ and\ \bibinfo {author} {\bibfnamefont
  {V.}~\bibnamefont {Awana}},\ }\bibfield  {title} {\bibinfo {title} {Crystal
  growth and basic transport and magnetic properties of mnbi 2 te 4},\
  }\href@noop {} {\bibfield  {journal} {\bibinfo  {journal} {Journal of
  Superconductivity and Novel Magnetism}\ }\textbf {\bibinfo {volume} {32}},\
  \bibinfo {pages} {3705} (\bibinfo {year} {2019})}\BibitemShut {NoStop}%
\bibitem [{\citenamefont {Jain}\ \emph {et~al.}(2013)\citenamefont {Jain},
  \citenamefont {Ong}, \citenamefont {Hautier}, \citenamefont {Chen},
  \citenamefont {Richards}, \citenamefont {Dacek}, \citenamefont {Cholia},
  \citenamefont {Gunter}, \citenamefont {Skinner}, \citenamefont {Ceder},\ and\
  \citenamefont {Persson}}]{project}%
  \BibitemOpen
  \bibfield  {author} {\bibinfo {author} {\bibfnamefont {A.}~\bibnamefont
  {Jain}}, \bibinfo {author} {\bibfnamefont {S.~P.}\ \bibnamefont {Ong}},
  \bibinfo {author} {\bibfnamefont {G.}~\bibnamefont {Hautier}}, \bibinfo
  {author} {\bibfnamefont {W.}~\bibnamefont {Chen}}, \bibinfo {author}
  {\bibfnamefont {W.~D.}\ \bibnamefont {Richards}}, \bibinfo {author}
  {\bibfnamefont {S.}~\bibnamefont {Dacek}}, \bibinfo {author} {\bibfnamefont
  {S.}~\bibnamefont {Cholia}}, \bibinfo {author} {\bibfnamefont
  {D.}~\bibnamefont {Gunter}}, \bibinfo {author} {\bibfnamefont
  {D.}~\bibnamefont {Skinner}}, \bibinfo {author} {\bibfnamefont
  {G.}~\bibnamefont {Ceder}},\ and\ \bibinfo {author} {\bibfnamefont {K.~A.}\
  \bibnamefont {Persson}},\ }\bibfield  {title} {\bibinfo {title} {Commentary:
  The materials project: A materials genome approach to accelerating materials
  innovation},\ }\href {https://doi.org/10.1063/1.4812323} {\bibfield
  {journal} {\bibinfo  {journal} {APL Materials}\ }\textbf {\bibinfo {volume}
  {1}},\ \bibinfo {pages} {011002} (\bibinfo {year} {2013})},\ \Eprint
  {https://arxiv.org/abs/https://pubs.aip.org/aip/apm/article-pdf/doi/10.1063/1.4812323/13163869/011002\_1\_online.pdf}
  {https://pubs.aip.org/aip/apm/article-pdf/doi/10.1063/1.4812323/13163869/011002\_1\_online.pdf}
  \BibitemShut {NoStop}%
\end{thebibliography}%

\end{document}